\documentclass[aps,twocolumn,pra,superscriptaddress,amsmath,showpacs,tightenlines,nofootinbib]{revtex4}
\usepackage{epsfig,graphicx,times}
\usepackage{amstext}
\usepackage{amsmath}            
\usepackage{amssymb}            
\usepackage{graphicx}           
\usepackage{latexsym}
\usepackage{bm}
\usepackage{color}
\begin{document}
\title{Controllable optical response by modifying the gain and loss of a mechanical resonator

and cavity mode in an optomechanical system}
\author{Yu-Long Liu}
\affiliation{Institute of Microelectronics, Tsinghua University, Beijing 100084, China}
\affiliation{Tsinghua National Laboratory for Information Science and Technology (TNList), Beijing 100084, China}
\author{Rebing Wu}
\email{rbwu@tsinghua.edu.cn}
\affiliation{Tsinghua National Laboratory for Information Science and Technology (TNList), Beijing 100084, China}
\affiliation{Department of Automation, Tsinghua University, Beijing 100084, China}
\author{Jing Zhang}
\affiliation{Tsinghua National Laboratory for Information Science and Technology (TNList), Beijing 100084, China}
\affiliation{Department of Automation, Tsinghua University, Beijing 100084, China}
\author{\c{S}ahin Kaya \"{O}zdemir}
\affiliation{Department of Electrical and Systems Engineering, Washington University, St.~Louis, Missouri 63130, USA}
\author{Lan Yang}
\affiliation{Department of Electrical and Systems Engineering, Washington University, St.~Louis, Missouri 63130, USA}
\author{Franco Nori}
\affiliation{CEMS, RIKEN, Saitama 351-0198, Japan}
\affiliation{Department of Physics, The University of Michigan, Ann Arbor, Michigan 48109-1040, USA}
\author{Yu-xi Liu}
\email{yuxiliu@mail.tsinghua.edu.cn}
\affiliation{Institute of Microelectronics, Tsinghua University, Beijing 100084, China}
\affiliation{Tsinghua National Laboratory for Information Science and Technology (TNList), Beijing 100084, China}
\date{\today }

\begin{abstract}
We theoretically study a strongly-driven optomechanical system which consists of a passive optical cavity and an active mechanical resonator. When the optomechanical coupling strength is varied, phase transitions, which are similar those observed in $\mathcal{PT}$-symmetric systems, are observed. We show that the optical transmission can be controlled by changing the gain of the mechanical resonator and loss of the optical cavity mode. Especially, we find that: (i) for balanced gain and loss, optical amplification and absorption can be tuned by changing the optomechanical coupling strength through a control field; (ii) for unbalanced gain and loss, even with a tiny mechanical gain, both optomechanically-induced transparency and anomalous dispersion can be observed around a critical point, which exhibits an ultra-long group delay. The time delay $\tau$ can be optimized by regulating the optomechanical coupling strength through the control field and improved up to several orders of magnitude ($\tau\sim2$ $\mathrm{ms}$) compared to that of conventional optomechanical systems ($\tau\sim1$ $\mu\mathrm{s}$). The presence of mechanical gain makes the group delay more robust to
environmental perturbations. Our proposal provides a powerful platform to control light transport using a $\mathcal{PT}$-symmetric-like optomechanical system.
\end{abstract}

\maketitle

\pagenumbering{arabic}

\section{Introduction}

Optomechanical systems explore the radiation-pressure induced interaction between electromagnetic fields and mechanical systems~\cite{kippenberg,marquart,aspelmeyer}. These provide one of the most promising platforms to study and understand quantum-mechanical laws on a macroscopic scale. Many breakthroughs along this research direction have emerged in recent years~\cite{Revmodphys}. In particular, the demonstrations of optomechanically-induced transparency (OMIT)~\cite{peit1,peit3,peit5}, absorption~\cite{peit4}, and amplification (OMIA)~\cite{peit6,microwaveamp}, allow the control of light propagation at room temperature by using nano- and micro-fabricated optomechanical structures.

Recently, the experimental realization of parity-time ($\mathcal{PT}$)-symmetric Hamiltonian systems have attracted extensive attention~\cite{ptf,pth,pti,ptj,ptk,ptm}. Although being originally explored at a highly mathematical level~\cite{bender}, $\mathcal{PT}$-symmetric systems have been realized in open physical structures, especially in optical devices that have balanced optical loss and gain~\cite{bender2}.

It is known that the eigenvalues of a $\mathcal{PT}$-symmetric Hamiltonian may be real, and a phase transition to broken $\mathcal{PT}$-phase may occur when the $\mathcal{PT}$-symmetry condition is broken. Across the transition point (i.e., in the broken-$\mathcal{PT}$ phase), pairs of eigenvalues coalesce and become conjugate complex numbers. Phase transitions have been observed in waveguides~\cite{pta,ptb,ptd,pte}, and microcavities~\cite{PT1,PT2}. Around the $\mathcal{PT}$-phase transition point, many unique optical phenomena (such as loss-induced transparency~\cite{pta}, power oscillations violating left-right symmetry~\cite{ptb}, low-power optical diodes~\cite{PT1}, and single-mode laser~\cite{singlemodeone,singlemodetwo}) have been demonstrated.

Very recently, the effects of $\mathcal{PT}$-symmetry and its breaking in coupled-cavity optomechanics have been explored. Such $\mathcal{PT}$-symmetric structures are realized by coupling an active (with optical gain) optical cavity with no mechanical mode to a passive (lossy) optical cavity supporting a mechanical mode. Benefiting from the phase transition, ultra-low threshold optical chaos~\cite{ptchaos}, phonon laser~\cite{ptphonon}, and inverted OMIT~\cite{pteit} have been theoretically proposed. However, for coupled-cavity $\mathcal{PT}$-symmetric systems, it is not easy to achieve high optical gain in the active cavity to balance the loss of the passive optical cavity. Moreover, a tunable coupling between the active and passive optical cavities without moving the cavities is desirable to induce a phase transition for a given gain-to-loss ratio.

Due to the fact that nontraditional control of optical fields or mechanical oscillations is always associated with the phase transition in a
$\mathcal{PT}$-symmetric system, a natural question is whether the phenomena occurring in previously studied $\mathcal{PT}$-symmetric devices can be demonstrated in a single standard optomechanical system by modifying the gain and loss of the mechanical resonator and the optical cavity mode, such that it can also exhibit a phase transition.

Thanks to recent progress in the coherent manipulation of phonons~\cite{phonon,phonon1,phonon2,phonon3,phonon4,phononpump1,phononpump2,phononpump3} (e.g., phonon laser, phonon pump), it is now clear that a considerable gain can be introduced to a mechanical resonator. With an additional optical field serving as the control, a controllable optomechanical coupling with strength $G$ can be achieved between the lossy optical cavity and the active (i.e., with mechanical gain) mechanical resonator. Here we show that the mechanical gain and the controllable coupling enable the construction of a $\mathcal{PT}$-symmetric-like optomechanical system. By modifying the gain of the mechanical resonator and the loss of the optical cavity mode, a phase transition similar to the one in $\mathcal{PT}$-symmetric systems can be observed by changing the effective optomechanical coupling strength. The photoelastic scattering and anti-Stokes field are greatly enhanced by the mechanical gain. Such enhancement can be used to control the transmission of an optical probe field and improve the system robustness to environmental perturbations. Especially, we will theoretically show:

(i) For the balanced gain-loss case, tunable optical signal amplification and absorption can be achieved. The transition from optical absorption (amplification) to amplification (absorption) is controlled by increasing (decreasing) the optomechanical coupling strength $G$.

(ii) For the unbalanced gain-loss case (even with a tiny mechanical gain), a critical point is found in the broken $\mathcal{PT}$-symmetry regime. Ultra-slow light appears around this critical point.

(iii) With the help of the mechanical gain, the group delay becomes much more robust to the fiber-cavity coupling perturbation. This can help to build stable optical delay lines.

These theoretical findings enable new applications for controlling the transmission of light beyond what is possible with conventional optomechanical systems~\cite{peit1,peit3,peit5,peit4,peit6,microwaveamp}.

The remainder of the paper is organized as follows. In Sec.~II we introduce the theoretical model of the $\mathcal{PT}$-symmetric-like optomechanical system. In Sec.~III we study the linearized and stable optomechanical coupling which is used to control the phase transitions in the $\mathcal{PT}$-symmetric-like optomechanical system. In Sec.~IV we discuss the effects of the mechanical gain and $\mathcal{PT}$-symmetry-like structure on the control of optical transmission. The unique control of output fields, especially for
mechanical-gain-induced optical amplification, absorption, unconventional OMIT, ultra-slow light, and enhanced system robustness, is also demonstrated. Conclusions discussions are presented in Sec.~VI. Different ways to obtain the mechanical gain are discussed and shown
in Appendix A.

\begin{figure}[ptb]
\includegraphics[bb=58 25 605 521,  width=9cm, clip]{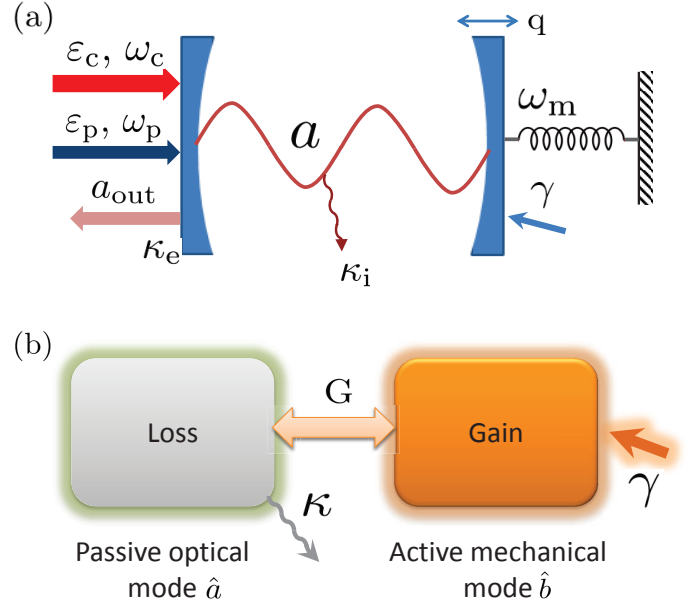}\caption{(Color online) (a) Schematic of a $\mathcal{PT}$-symmetric-like optomechanical system consisting of a passive optical cavity and an active mechanical resonator. The optical cavity is driven by a strong control field of frequency $\omega_{\mathrm{c}}$ and a weak probe field with frequency $\omega _{\mathrm{p}}$ from the left-hand side. The coupling strengths of the control and probe fields to the cavity modes are $\varepsilon_{\mathrm{c}}$ and $\varepsilon_{\mathrm{p}}$, respectively. $\kappa_{\mathrm{e}}$ and $\kappa_{\mathrm{i}}$ denote, respectively, the intrinsic and coupling loss of the optical cavity. $a$ is the intracavity field, $a_{\mathrm{out}}$ is the output field, and $q$ represents the displacement of the movable mirror which
forms the optical cavity. A mechanical gain $\gamma$ is introduced to the mechanical resonator supporting a mechanical mode with frequency $\omega_{\mathrm{m}}$. (b) The equivalent coupled-harmonic-resonators model with gain and loss for the $\mathcal{PT}$-symmetric optomechanical system. The passive cavity field with a total loss rate of $\kappa$ and the active mechanical mode with gain $\gamma$ are coupled with a controllable optomechanical coupling strength $G$.}%
\label{fig1ab}%
\end{figure}

\section{Model}

As schematically shown in Fig.~\ref{fig1ab}, we study an optomechanical system that consists of a passive optical cavity and an active mechanical resonator. The mechanical gain is introduced by some experimentally feasible methods (see the detailed discussions in Appendix A).
In the following we will show that this system exhibits features typical of $\mathcal{PT}$-symmetric systems~\cite{PT1,PT2}. Thus, we call it as a $\mathcal{PT}$-symmetric-like optomechanical system. For conventional optomechanical systems, however, the passive optical cavity only couples to a passive mechanical resonator. The cavity field is coupled to a strong control field and a weak probe field with frequencies $\omega_{\mathrm{c}}$ and $\omega_{\mathrm{p}}$, respectively. The coupling strength between the cavity field and the control
(probe) field is $\varepsilon_{\mathrm{c}}$ ($\varepsilon_{\mathrm{p}}$). In the rotating reference frame at the frequency $\omega_{\mathrm{c}}$ of the control field, the Hamiltonian of the composite system can be written as%
\begin{align}
H  &  =\hbar\Delta_{\mathrm{a}}a^{\dag}a+\hbar\omega_{\mathrm{m}}b^{\dag
}b-\hbar g_{\mathrm{0}}a^{\dag}a(b^{\dag}+b)\nonumber\\
&  +i\hbar(\varepsilon_{\mathrm{c}}a^{\dag}+\varepsilon_{\mathrm{p}%
}e^{-i\delta t}a^{\dag}-\mathrm{H.c}.), \label{Ha}%
\end{align}
where $a$ ($a^{\dag}$) is the annihilation (creation) operator of the cavity field with frequency $\omega_{\mathrm{a}}$ in the absence of the mechanical resonator, $b$ ($b^{\dagger}$) is the annihilation (creation) operator of mechanical mode with resonance frequency $\omega_{\mathrm{m}}$, $\Delta_{\mathrm{a}}=\omega_{\mathrm{a}}-\omega_{\mathrm{c}}$ ($\delta=\omega_{\mathrm{p}}-\omega_{\mathrm{c}}$) is the frequency detuning between the cavity (probe) field and the control field, and $g_{\mathrm{0}}$ is the single-photon optomechanical coupling strength. The amplitudes of the control and the probe fields are normalized to the photon flux at the input of the cavity, i.e., $\varepsilon_{\mathrm{c}}=\sqrt{P_{\mathrm{c}}\kappa_{\mathrm{e}}/\hbar\omega_{\mathrm{c}}}$ and $\varepsilon_{\mathrm{p}}%
=\sqrt{P_{\mathrm{p}}\kappa_{\mathrm{e}}/\hbar\omega_{\mathrm{p}}}$, where $P_{\mathrm{c}}$ ($P_{\mathrm{p}}$) is the power of the control (probe) field, and $\kappa_{\mathrm{e}}$ is the the external loss rate. Without loss of generality, we assume the total loss of the cavity is $\kappa$, which is the sum of the intrinsic loss rate $\kappa_{\mathrm{i}}$\ and the external loss rate $\kappa_{\mathrm{e}}$. The coupling strength between external field and the cavity is defined as $\eta=\kappa_{\mathrm{e}}/(\kappa_{\mathrm{i}}+\kappa_{\mathrm{e}})$, which can be continuously adjusted in experiments. When the coupling parameter $\eta\ll1$, the system is in the undercoupling regime, when $\eta\simeq1$, the system is in the overcoupling regime, and when $\eta=1/2$, the system is in the critical coupling regime~\cite{coupling1,coupling2}.

The quantum Langevin equations for the operators $a$ and $b$ of the cavity field and mechanical mode are given by%
\begin{align}
\dot{a}  &  =-\left(  i\Delta_{\mathrm{a}}+\frac{\kappa}{2}\right)
a+ig_{\mathrm{0}}a\left(  b^{\dag}+b\right)  +\varepsilon_{\mathrm{c}%
}+\varepsilon_{\mathrm{p}}e^{-i\delta t}+f,\label{La}\\
\dot{b}  &  =-\left(  i\omega_{\mathrm{m}}-\frac{\gamma}{2}\right)
b+ig_{\mathrm{0}}a^{\dag}a+\xi, \label{Lb}%
\end{align}
where $\gamma$ represents the controllable gain of the mechanical resonator, while $f$ and $\xi$ are the quantum noise operators with zero-mean value, i.e., $\left\langle f\right\rangle =\left\langle \xi\right\rangle =0$. We are interested in the linear response of the driven optomechanical system to the weak probe field. So, under the condition $\varepsilon_{\mathrm{c}}%
\gg\varepsilon_{\mathrm{p}}$, Eqs.~(\ref{La}-\ref{Lb}) can be linearized by expanding the operators around their mean values, $a=\alpha+\delta
_{\mathrm{a}}$ and $b=$ $\beta+\delta_{\mathrm{b}}$:
\begin{align}
\dot{\delta}_{\mathrm{a}}  &  =-\left(  i\Delta+\frac{\kappa}{2}\right)
\delta_{\mathrm{a}}+iG(\delta_{\mathrm{b}}^{\dag}+\delta_{\mathrm{b}%
})+\varepsilon_{\mathrm{p}}e^{-i\delta t}+f,\label{Deltaaa}\\
\dot{\delta}_{\mathrm{b}}  &  =-\left(  i\omega_{\mathrm{m}}-\frac{\gamma}%
{2}\right)  \delta_{\mathrm{b}}+i(G\delta_{\mathrm{a}}^{\dag}+G^{\ast}%
\delta_{\mathrm{a}})+\xi, \label{Deltabb}%
\end{align}
where $G=g_{\mathrm{0}}\alpha$ is the effective optomechanical coupling strength, and $\Delta=\Delta_{\mathrm{a}}-g_{\mathrm{0}}(\beta^{\ast}+\beta)$ is the effective detuning between the cavity field and the control field. The mean values $\alpha$ and $\beta$ in the steady state can be calculated as%
\begin{align}
\alpha &  =\frac{\varepsilon_{\mathrm{c}}}{i\Delta+\frac{\kappa}{2}%
},\label{Alpha}\\
\beta &  =\frac{ig_{\mathrm{0}}\left\vert \alpha\right\vert ^{2}}%
{i\omega_{\mathrm{m}}-\frac{\gamma}{2}}. \label{Belta}%
\end{align}

We now assume that the optical cavity is driven in the red-sideband regime (e.g., $\Delta=\omega_{\mathrm{m}}$). Under the resolved sideband limit with $\omega_{\mathrm{m}}\gg(\kappa,\gamma)$, the rotating-wave approximation~\cite{peit1,peit3,peit4,peit5,peit6} can be applied for
Eqs.~(\ref{Deltaaa}-\ref{Deltabb}), which leads to%
\begin{align}
\dot{\delta}_{\mathrm{a}}  &  =-\left(  i\Delta+\frac{\kappa}{2}\right)
\delta_{\mathrm{a}}+iG\delta_{\mathrm{b}}+\varepsilon_{\mathrm{p}}e^{-i\delta
t}+f,\label{lineard}\\
\dot{\delta}_{\mathrm{b}}  &  =-\left(  i\omega_{\mathrm{m}}-\frac{\gamma}%
{2}\right)  \delta_{\mathrm{b}}+iG^{\ast}\delta_{\mathrm{a}}+\xi.
\label{lineardb}%
\end{align}
For simplicity, we move into another interaction picture by introducing $\delta_{\mathrm{a}}=Ae^{-i\delta t}$, $\delta_{\mathrm{b}}=Be^{-i\delta t}$, $f\rightarrow fe^{-i\delta t}$, and $\xi\rightarrow\xi e^{-i\delta t}$, so Eqs.~(\ref{lineard}-\ref{lineardb}) become%
\begin{align}
\dot{A}  &  =\left(  -i\omega_{1}-\frac{\kappa}{2}\right)  A+iGB+\varepsilon
_{\mathrm{p}}+f,\label{AA}\\
\dot{B}  &  =\left(  -i\omega_{2}+\frac{\gamma}{2}\right)  B+iG^{\ast}A+\xi,
\label{BB}%
\end{align}
where $\omega_{1}=\Delta-\delta$, and $\omega_{2}=\omega_{\mathrm{m}}-\delta$.

Then, we take the expectation values of the operators in Eqs.~(\ref{AA}-\ref{BB}). Note that the mean values of the quantum noise terms are zero, e.g., $\left\langle f\right\rangle =\left\langle \xi\right\rangle =0$. Under the steady-state condition $\left\langle \dot{A}\right\rangle =\left\langle\dot{B}\right\rangle =0$, one has%
\begin{align}
\left\langle A\right\rangle  &  =\frac{\left(  i\omega_{2}-\frac{\gamma}%
{2}\right)  \varepsilon_{\mathrm{p}}}{\left(  i\omega_{1}+\frac{\kappa}%
{2}\right)  \left(  i\omega_{2}-\frac{\gamma}{2}\right)  +\left\vert
G\right\vert ^{2}},\label{EAA}\\
\left\langle B\right\rangle  &  =\frac{iG^{\ast}\varepsilon_{\mathrm{p}}%
}{\left(  i\omega_{1}+\frac{\kappa}{2}\right)  \left(  i\omega_{2}%
-\frac{\gamma}{2}\right)  +\left\vert G\right\vert ^{2}}. \label{EBB}%
\end{align}
From Eqs.~(\ref{AA}-\ref{BB}), we can see that if the optical cavity is driven by a strong control field, the interaction between cavity field and mechanical resonator can be linearized, so that the system can be treated as two coupled harmonic oscillators with frequencies $\omega_{\mathrm{1}}$ and $\omega_{\mathrm{2}}$, respectively. The resulting effective Hamiltonian of the linearized $\mathcal{PT}$-symmetric-like optomechanical system is%
\begin{align}
H_{\mathrm{eff}}  &  =\hbar\omega_{\mathrm{1}}A^{\dag}A+\hbar\omega
_{\mathrm{2}}B^{\dag}B-\hbar\left(  GA^{\dag}B+G^{\ast}B^{\dag}A\right)
\nonumber\\
&  +\hbar\varepsilon_{\mathrm{p}}\left(  A+A^{\dag}\right)  . \label{Heff}%
\end{align}
Taking the cavity loss $\kappa$ and mechanical gain $\gamma$ into consideration, we obtain a non-Hermitian Hamiltonian as follows%
\begin{align}
H_{\mathrm{non}}  &  =\hbar\left(  \omega_{\mathrm{1}}-i\frac{\kappa}%
{2}\right)  A^{\dag}A+\hbar\left(  \omega_{\mathrm{2}}+i\frac{\gamma}%
{2}\right)  B^{\dag}B\nonumber\\
&  -\hbar\left(  GA^{\dag}B+G^{\ast}B^{\dag}A\right)  +\hbar\varepsilon
_{\mathrm{p}}\left(  A+A^{\dag}\right)  . \label{Hnon}%
\end{align}

The equivalent physical model corresponding to Eq.~(\ref{Hnon}) is shown in Fig.~\ref{fig1ab}(b). The passive cavity mode (photon) and active mechanical mode (phonon) are coupled, forming a $\mathcal{PT}$-symmetric physical system~\cite{PT1,PT2,bender,bender2,singlemodeone,singlemodetwo,ptchaos,ptphonon,pteit}. The effective coupling rate $G=g_{\mathrm{0}}\alpha$ between modes $A$ and $B$ can be continuously adjusted by tuning the power of the control field. Such controllable coupling parameter $G$ can be used to generate phase transitions similar to standard $\mathcal{PT}$-symmetric systems, which will be discussed in detail in the following section.

From the above discussions, we summarize the two key requirements for realizing such $\mathcal{PT}$-symmetric-like optomechanical system. One is to obtain mechanical gain under experimentally-feasible conditions, which will be discussed in appendix A. The other is to obtain an effective optomechanical coupling strength $G$ which is controlled by the power of the control field. Below we mainly discuss the behaviour of the phase transition in the $\mathcal{PT}$-symmetric-like optomechanical system studied here.

\section{Phase transition in optomechanical system by modifing gain and loss}

It is well known that $\mathcal{PT}$-symmetric systems can exhibit a phase transition (e.g., spontaneous $\mathcal{PT}$-symmetry breaking). The transition from the $\mathcal{PT}$-symmetry phase (real spectrum) to the broken $\mathcal{PT}$-symmetry (complex spectrum) phase occurs when a continuously adjusted parameter (controlling the degree of the system Hamiltonian non-Hermiticity) exceeds a critical value~\cite{PT1,PT2,bender}. In our system, such a control parameter is the effective optomechanical coupling strength $G$. The continuous adjustment of $G$ can be realized by changing the amplitude of the control field. Because optical bistability may occur when the control field is sufficiently strong~\cite{kbis,gong}, the following discussion will be restricted to cases with achievable effective optomechanical coupling $G$ before the system loses its stability.

\subsection{Effective optomechanical coupling}

From Eqs.~(\ref{Alpha}-\ref{Belta}), the mean intracavity photon number $I_{\mathrm{a}}=\left\vert \alpha\right\vert ^{2}$ in the steady state satisfies%
\begin{equation}
I_{\mathrm{a}}\left[  \frac{\kappa^{2}}{4}+\left(  \Delta_{\mathrm{a}}-2\chi
I_{\mathrm{a}}\right)  ^{2}\right]  =\left\vert \varepsilon_{\mathrm{c}%
}\right\vert ^{2}, \label{bistability}%
\end{equation}
where $\chi=g_{\mathrm{0}}^{2}/\omega_{\mathrm{m}}$. Since Eq.~(\ref{bistability}) is cubic in $I_{\mathrm{a}}$, the system may exhibit
bistability in a certain parameter regime. To see clearly when the system exhibits bistable behavior, we derive the bistability condition by imposing the condition that $\partial\left\vert \varepsilon_{\mathrm{c}}\right\vert^{2}/\partial I_{\mathrm{a}}=0$, which results in:%
\begin{equation}
\frac{\kappa^{2}}{4}+\Delta_{\mathrm{a}}^{2}-8\chi\Delta_{\mathrm{a}%
}I_{\mathrm{a}}+12\chi^{2}I_{\mathrm{a}}^{2}=0. \label{TWO}%
\end{equation}
The system becomes bistable when the discriminant of the above quadratic equation is positive, which gives%
\begin{equation}
4\Delta_{\mathrm{a}}^{2}-3\kappa^{2}>0.
\end{equation}
Since the cavity is driven in the red-sideband regime with $\Delta_{\mathrm{a}}=\omega_{\mathrm{m}}$, and under the resolved-sideband limit with $\omega_{\mathrm{m}}\gg\kappa$, we conclude that the system becomes bistable when the control field is sufficiently strong. Recall that $\varepsilon_{\mathrm{c}}=\sqrt{P_{\mathrm{c}}\kappa_{\mathrm{e}}/\hbar\omega_{\mathrm{c}}}$, and the threshold $P_{\mathrm{thr}}$ of bistability for the control optical field can be solved as:%
\begin{equation}
P_{\mathrm{thr}}=\frac{\hbar\omega_{\mathrm{c}}I_{\mathrm{a}}^{+}}%
{\kappa_{\mathrm{e}}}\left[  \frac{\kappa^{2}}{4}+\left(  \Delta_{\mathrm{a}%
}-2\chi I_{\mathrm{a}}^{+}\right)  ^{2}\right]  ,
\end{equation}
where%
\begin{equation}
I_{\mathrm{a}}^{+}=\frac{4\omega_{\mathrm{m}}+\sqrt{4\omega_{\mathrm{m}}%
^{2}-3\kappa^{2}}}{12\chi}.
\end{equation}

\begin{figure}[ptb]
\includegraphics[bb=50 230 516 638,  width=8cm, clip]{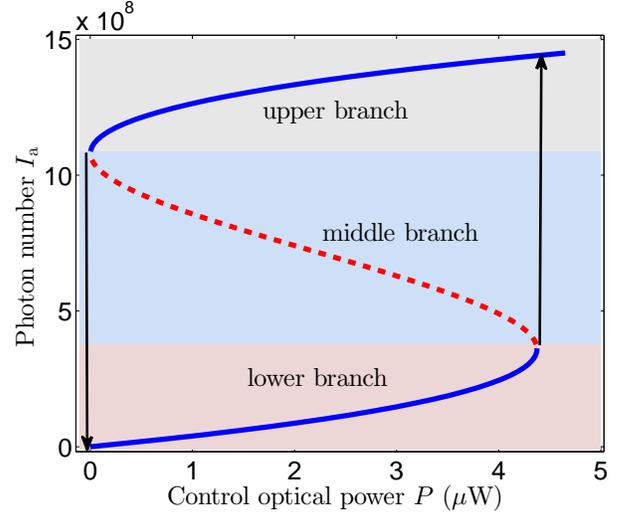}\caption{(Color online) Plot of the mean photons number $I_{\mathrm{a}}$ versus the power $P$ of the control optical pump ($\mu\mathrm{W}$) for a cavity-pump detuning of $\Delta_{\mathrm{a}}=\omega_{\mathrm{m}}$. Three different colors are used to represent the lower, middle, and upper branch, respectively. The other system
parameters are: $g_{0}/2\pi=230$ Hz, $\omega_{\mathrm{m}}/2\pi=10.69$ $\mathrm{MHz},$ $\kappa/2\pi=0.17$ $\mathrm{MHz}$ and $\eta=0.76$.}%
\label{fig2}%
\end{figure}

In Fig.~\ref{fig2}, the intracavity mean photon number $I_{\mathrm{a}}$ versus the input power $P$ of the control field is illustrated using experimentally realizable parameters~\cite{stcoup}. It is shown that the system exhibits optical bistability. The threshold of bistability for the control field is very low. In our case, for parameters given in Fig.~\ref{fig2}, $P_{\mathrm{thr}}=1.87$ nW. The bistability can be
observed by scanning the input pump power in two directions. For example, by gradually increasing the pump power from zero to a sufficiently strong pump power, about 4.3 $\mu$W, one finds the lower bistable point. The hysteresis then follows the arrow and jumps to the upper branch. The other unstable point can be obtained by gradually decreasing the input pump power to lower values, which appears at 1.87 nW. Note that the position of the bistable points strongly depends on the detuning $\Delta_{\mathrm{a}}$ between the cavity field and the control field. As indicated in Ref.~\cite{kbis}, the lower branch is always stable. So, in the following, we choose $P<4.3$ $\mu$W, which supports stable solutions.

\begin{figure}[ptb]
\includegraphics[bb=45 236 516 640,  width=8cm, clip]{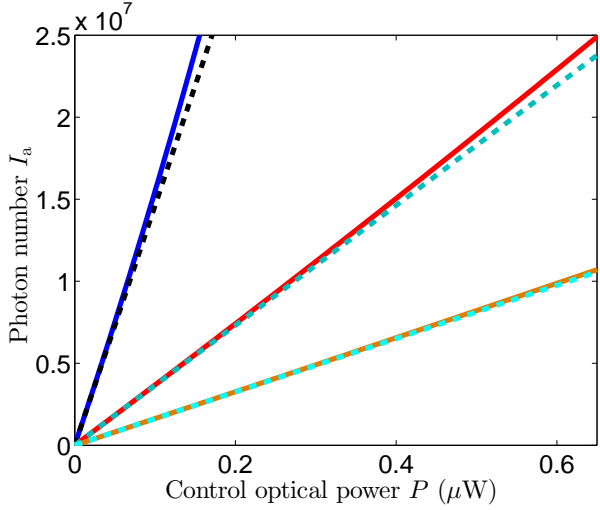}\caption{(Color online) Mean photon number $I_{\mathrm{a}}$ versus the power $P$ of the control optical pump ($\mu\mathrm{W}$) for different cavity-pump detunings ($\Delta_{\mathrm{a}}$). The solid curves represent the exact solutions calculated from Eq.~(\ref{bistability}), and the dashed curves represent the solutions of the approximating expression given in Eq.~(\ref{propor}). The detuning parameters are chosen as $\Delta_{\mathrm{a}}=0.5\omega_{\mathrm{m}}$ for the blue-solid and black-dashed curves, $\Delta_{\mathrm{a}}=\omega_{\mathrm{m}}$ for the red-solid and green-dashed curves and $\Delta_{\mathrm{a}}=1.5\omega_{\mathrm{m}}$ for the yellow-solid and cyan-dashed curves. The other system parameters used here are the same as those used in Fig.~\ref{fig2}.}%
\label{fig3}%
\end{figure}

We now consider the case when the system is driven by a relatively low pump power. As shown in Fig.~\ref{fig3}, the pump power $P$\ is chosen to be below $650$ nW, in which regime, we have%
\begin{equation}
I_{\mathrm{a}}\approx\frac{4\kappa_{\mathrm{e}}}{\hbar\omega_{\mathrm{c}}%
}\frac{P}{\kappa^{2}+4\Delta_{\mathrm{a}}^{2}}. \label{propor}%
\end{equation}
That is, when the pump power is low, the intracavity photon number is approximately proportional to the pump power. Figure~\ref{fig3} depicts the relation between the intracavity photon number and the pump power using Eqs.~(\ref{bistability}) and (\ref{propor}), respectively. These two results agree well. It is easy to find that under a fixed pump power, the intracavity photon number goes up when the cavity-pump laser detuning $\Delta_{\mathrm{a}}$ decreases.

\begin{figure}[ptb]
\includegraphics[bb=45 236 516 640,  width=8cm, clip]{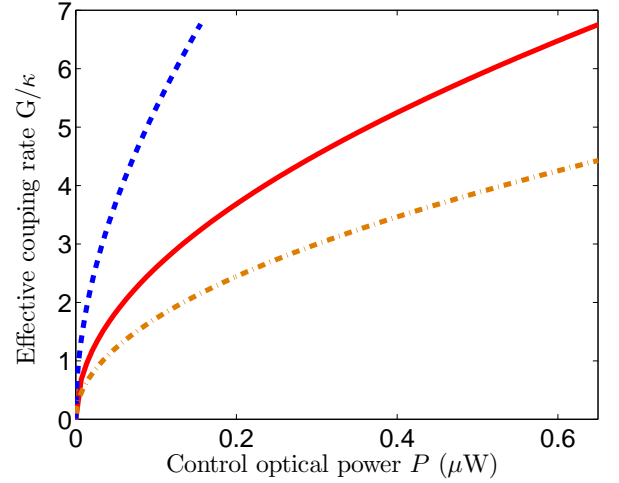}\caption{(Color online) Effective optomechanical coupling strength $G$ versus the control optical power $P$ ($\mu\mathrm{W}$) for different cavity-pump detunings ($\Delta_{\mathrm{a}}$). The detuning parameters are chosen as $\Delta_{\mathrm{a}}=0.5\omega_{\mathrm{m}}$ for the blue-dashed line, $\Delta_{\mathrm{a}}=\omega_{\mathrm{m}}$ for the red-solid curve, and $\Delta_{\mathrm{a}}=1.5\omega_{\mathrm{m}}$ for the orange-dashed-dotted curve. The other system parameters used here are the same as in Fig.~\ref{fig2}.}%
\label{fig4}%
\end{figure}

The dependence of the effective optomechanical coupling strength\ $G=g_{\mathrm{0}}\alpha$ on the pump power $P$ is shown in Fig.~\ref{fig4}. It is seen that $G$ can be much larger than the cavity field dissipation rate $\kappa$. Thus, a strong control field can induce a strong coupling $G$ between the cavity field and the mechanical mode. In addition, due to the dynamical backaction, the control field will also affect the mechanical dissipation rate. Nevertheless, it can be compensated by the controllable mechanical gain introduced by the control field as well.

\subsection{Coupling strength controlled phase transition}

A $\mathcal{PT}$-symmetric system can experience a phase transition, which has been shown to lead to the observations of interesting and counterintuitive phenomena~\cite{PT1,PT2,bender,bender2,singlemodeone,singlemodetwo,ptchaos,ptphonon,pteit}. We now discuss how phase transition can take place in a $\mathcal{PT}$-symmetric-like optomechanical system. From Eq.~(\ref{Hnon}), we can obtain an effective non-Hermitian Hamiltonian as%
\begin{align}
\tilde{H}_{\mathrm{eff}}  &  =\hbar\left(  \omega_{\mathrm{1}}-i\frac{\kappa
}{2}\right)  A^{\dag}A+\hbar\left(  \omega_{\mathrm{2}}+i\frac{\gamma}%
{2}\right)  B^{\dag}B\nonumber\\
&  -\hbar\left(  GA^{\dag}B+G^{\ast}B^{\dag}A\right),
\end{align}
in which the weak probe field is omitted. The coupling of these two resonators leads to two supermodes, $A_{+}=(A+B)/\sqrt{2}$ and $A_{-}=(A-B)/\sqrt{2}$, with the eigenfrequencies $\omega_{+}$ and $\omega_{-}$ given by%
\begin{align}
\omega_{\pm}  &  =\frac{1}{2}(\omega_{\mathrm{1}}+\omega_{\mathrm{2}}%
)-\frac{i}{4}(\kappa-\gamma)\nonumber\\
&  \pm\sqrt{G^{2}-\left[  -\frac{i}{2}(\omega_{\mathrm{1}}-\omega_{\mathrm{2}%
})-\frac{\kappa+\gamma}{4}\right]  ^{2}}, \label{PTG}%
\end{align}
Experimentally, we can tune the resonances of the resonators to be degenerate, i.e., $\omega_{\mathrm{1}}=\omega_{\mathrm{2}}=\omega_{\mathrm{o}}$, in which case the eigenfrequencies in Eq.~(\ref{PTG}) of the supermodes become%
\begin{equation}
\omega_{\pm}=\omega_{\mathrm{o}}-\frac{i}{4}(\kappa-\gamma)\pm\sqrt
{G^{2}-\left(  \frac{\kappa+\gamma}{4}\right)  ^{2}}. \label{PTF}%
\end{equation}
where the expression in the square-root quantifies the competition between the coupling strength and the loss/gain of the resonators. The real (imaginary) part of $\omega_{\pm}$ quantify the frequencies (dissipation) of these two supermodes.

\begin{figure*}[ptb]
\includegraphics[bb=30 210 530 620,  width=0.4\textwidth, clip]{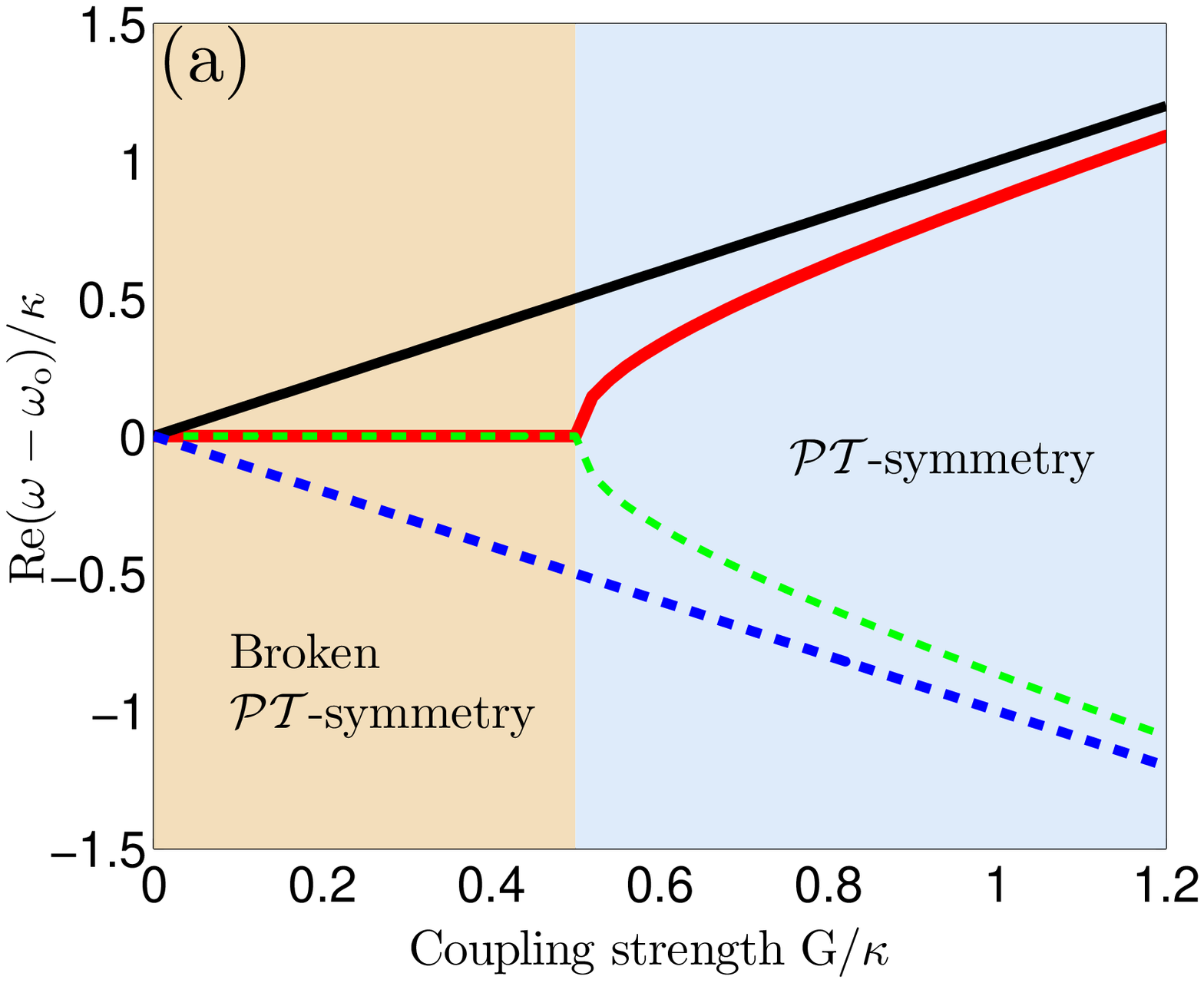}\includegraphics[bb=30 210 530 620,  width=0.4\textwidth, clip]{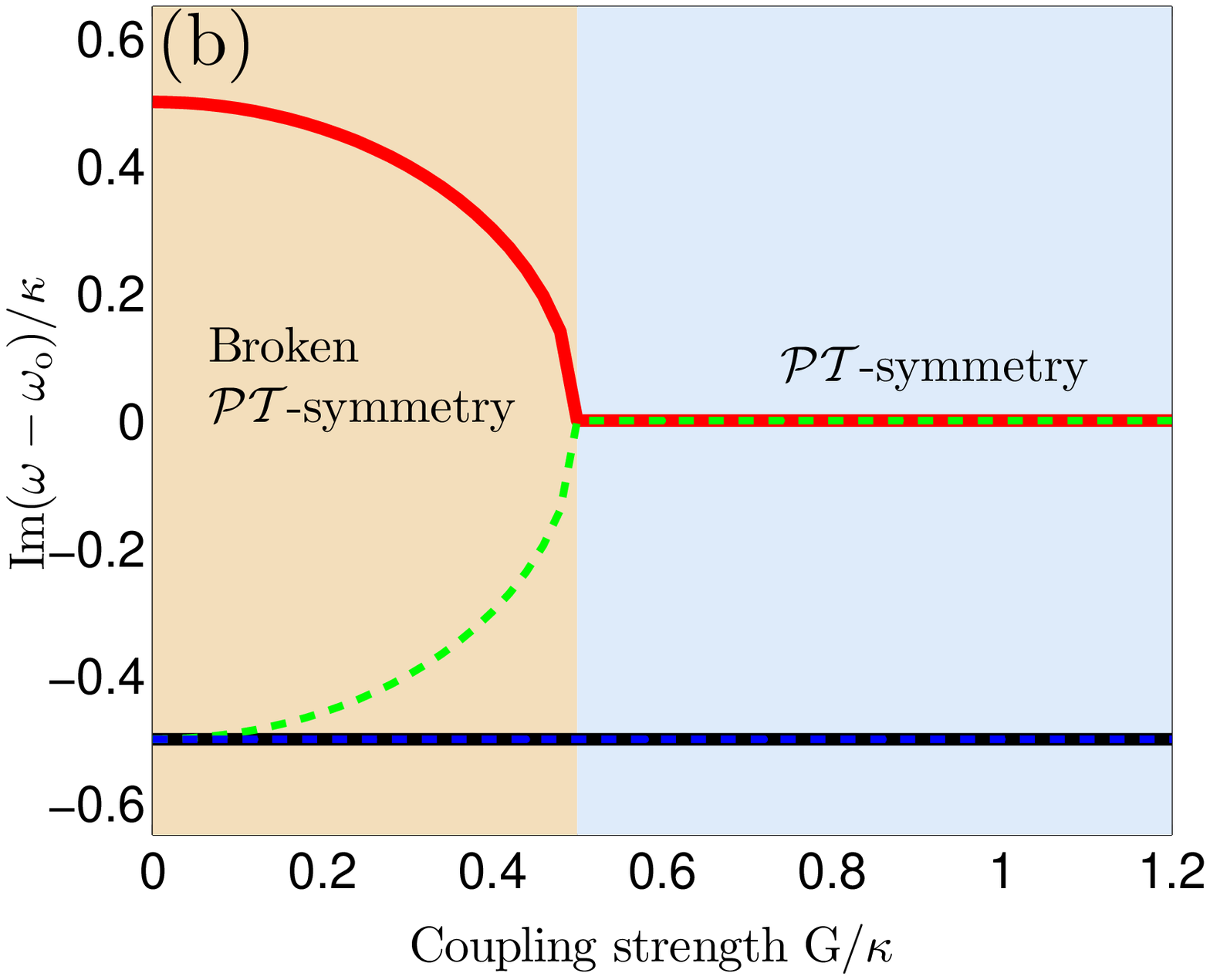}
\includegraphics[bb=30 210 530 620,  width=0.4\textwidth, clip]{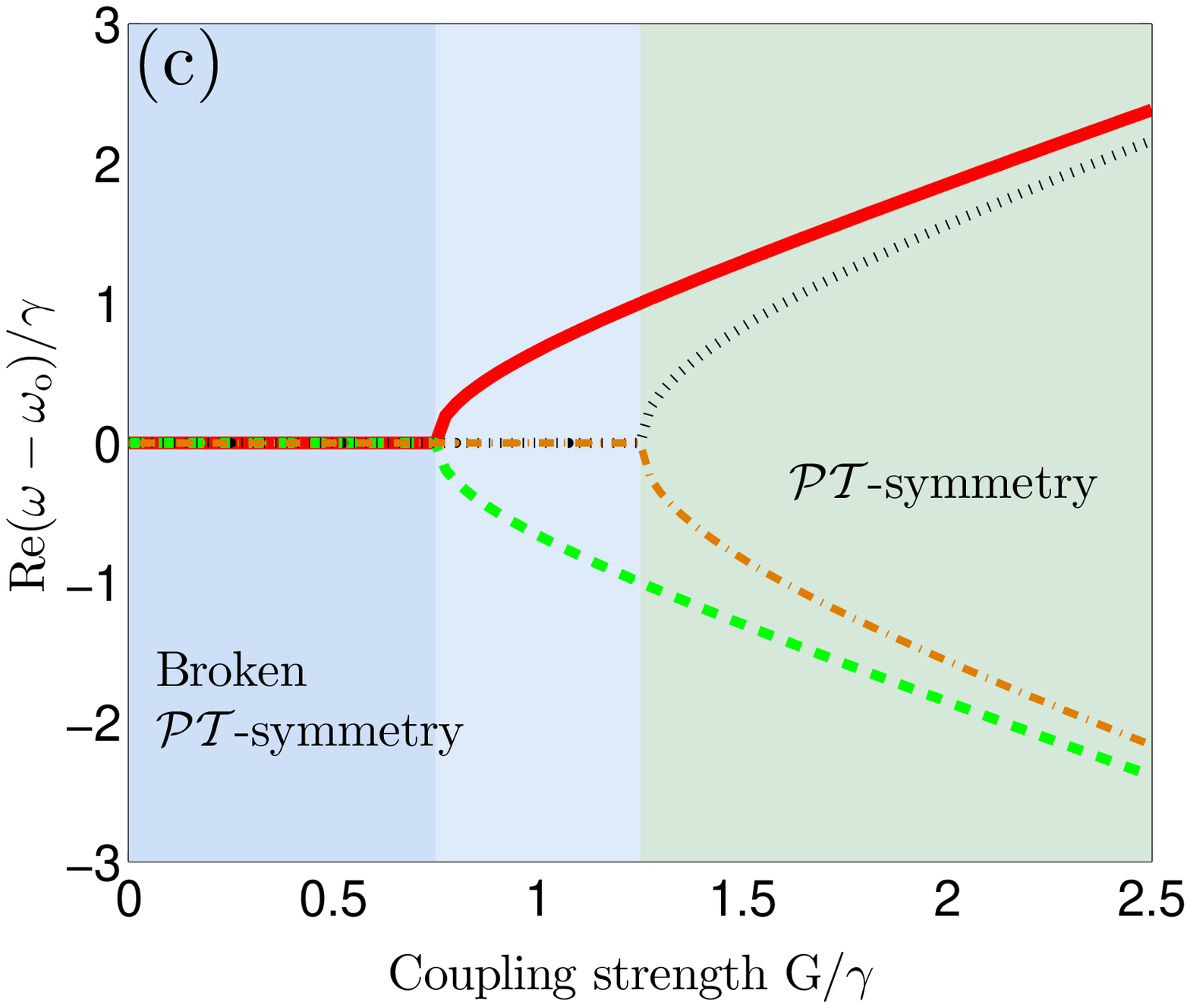}\includegraphics[bb=30 210 530 620,  width=0.4\textwidth, clip]{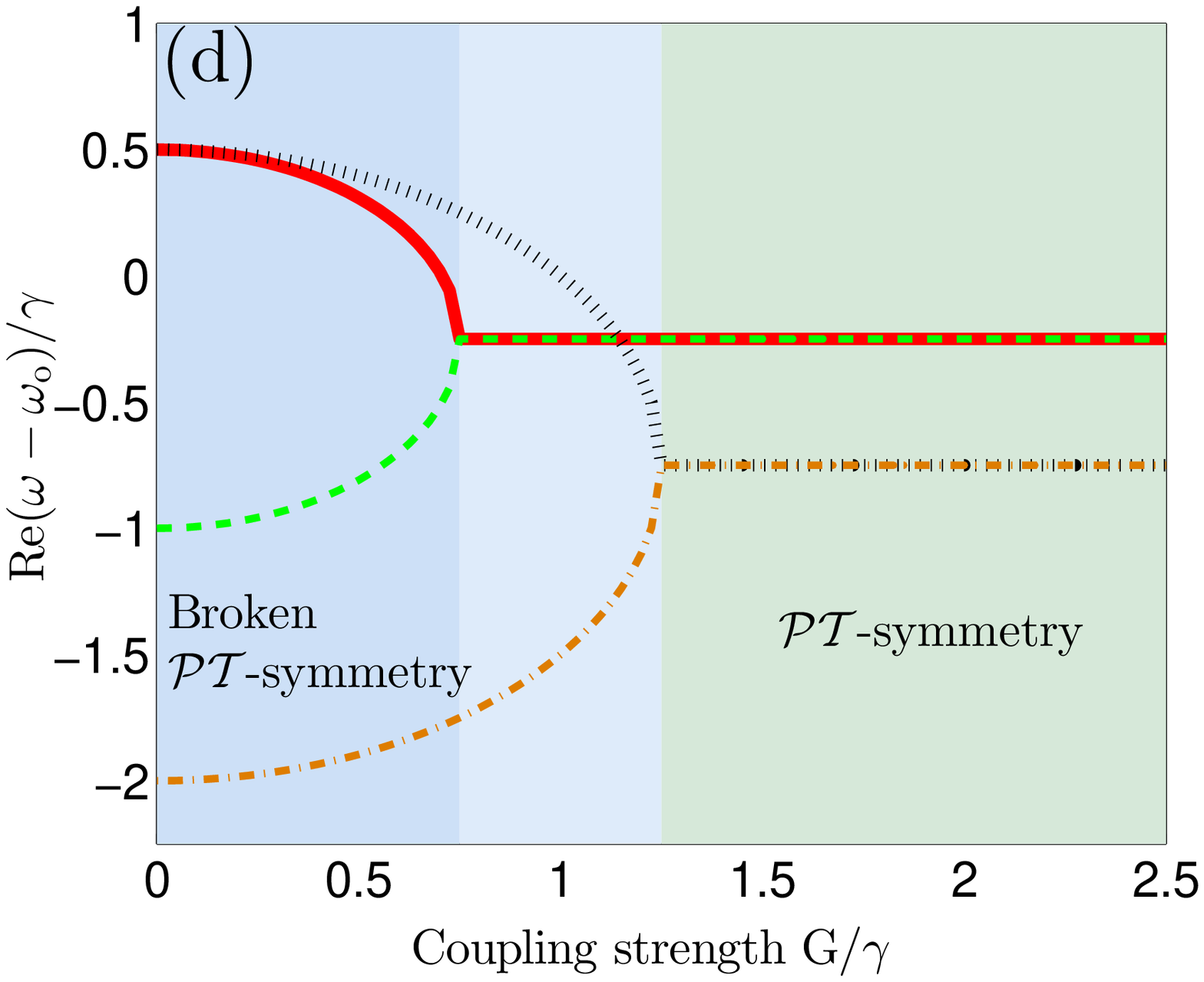}\caption{(Color online) The real and imaginary parts of eigenfrequencies $\omega_{\pm}$ versus coupling strength $G$ are plotted in (a) and (b), respectively. The red-solid and cyan-dashed curves in (a) and (b) represent the case with balanced gain and loss. The black-solid and blue-dashed curves in (a) and (b) represent the case without mechanical gain. The real and imaginary parts of eigenfrequencies $\omega_{\pm}$ for unbalanced gain-to-loss case are plotted in (c) and (d), respectively. The red-solid and cyan-dashed curves are calculated with $\kappa=2\gamma$. The black-dotted and yellow-dash-dotted curves are calculated with $\kappa=4\gamma$. Different background-colors are used to represent the broken and unbroken $\mathcal{PT}$-symmetric regimes.}%
\label{fig5}%
\end{figure*}

The real and imaginary parts of eigenfrequencies $\omega_{\pm}$ versus coupling strength $G$ are plotted in Fig.~\ref{fig5}(a) and (b), respectively. We first consider the case in which the gain and loss are strictly balanced [see red-solid and cyan-dashed curves in Figure~\ref{fig5}(a) and (b)]. There are two different regimes: (i) the strong coupling regime when $G>(\kappa+\gamma)/4$ (similarly, $G>\kappa/2$ or $G>\gamma/2$, when setting $\kappa=\gamma$), and (ii) the weak coupling regime when $G<(\kappa+\gamma)/4$ (similarly, $G<\kappa/2$ or $G<\gamma/2$, when setting $\kappa=\gamma$), which correspond to the $\mathcal{PT}$-symmetric regime and the broken $\mathcal{PT}$-symmetric regime, respectively, as in standard $\mathcal{PT}$-systems.

In the $\mathcal{PT}$-symmetric regime, the coupling between the optical resonator and the mechanical resonator creates two separate supermodes with different frequencies $\omega_{\mathrm{o}}\pm\sqrt{G^{2}-\left[  \left(\kappa+\gamma\right)  ^{2}/16\right]  }$, respectively. But these two supermodes have identical linewidths $(\kappa-\gamma)/4$, which reduces to zero-linewidth at exact balance of gain and loss (i.e., $\kappa=\gamma$).

In the broken $\mathcal{PT}$-symmetric regime, these two supermodes are degenerate with same frequency $\omega_{\mathrm{0}}$ but different linewidths $-(\kappa-\gamma)/4\pm\sqrt{\left[  \left(  \kappa+\gamma\right)^{2}/16\right]  -G^{2}}$, which reduces to $\pm\sqrt{\kappa^{2}/4-G^{2}}$ or $\pm\sqrt{\gamma^{2}/4-G^{2}}$ (i.e., exact gain and loss balance).

At the critical point $G=(\kappa+\gamma)/4$, both the resonance frequencies and the linewidths of these two supermodes coincide. This critical point is the $\mathcal{PT}$ phase transition point or the exceptional point (EP) ~\cite{PT1,PT2,bender,bender2,singlemodeone,singlemodetwo,ptchaos,ptphonon,pteit}. As shown in Fig.~\ref{fig5}(a) and Fig.~\ref{fig5}(b), by increasing the effective optomechanical coupling rate $G$, one can realize the transition from the broken to the unbroken $\mathcal{PT}$-symmetric phase, and the phase transition point occurs at $G=(\kappa+\gamma)/4$.

In contrast, if both the cavity field and the mechanical mode are passive and have the same loss $\kappa$, the supermodes always have different frequencies $\omega_{\mathrm{o}}\pm G$ but identical linewidth $\kappa/2$ [see black-solid and blue-dashed curves in Figure~\ref{fig5}(a) and (b)]. Also, the frequency separation is proportional to the coupling strength $G$, and no phase transition will be observed for this passive-passive coupling case.

We now consider the general case in which the gain and loss are not exactly balanced. Two examples with $r=2$ and $r=4$ are shown in Fig.~\ref{fig5}(c) and Fig.~\ref{fig5}(d), respectively. The critical coupling strength is given by $G_{\mathrm{c}}=\gamma(1+r)/4$, which implies that a stronger coupling strength $G$ is required for a $\mathcal{PT}$ phase transition for the case of a higher loss-gain ratio $r$. In contrast with Fig.~\ref{fig5}(a) and Fig.~\ref{fig5}(b), the frequencies $\omega_{\pm}$ cannot be purely real in the unbalanced case. When the coupling is weak, i.e., $G<(\kappa+\gamma)/4,$ two degenerated supermodes appear with different linewidths (one with gain and the other with loss). When the coupling is strong enough, $G>(\kappa+\gamma)/4$, the supermodes exhibit different frequencies but identical linewidths.

\section{Controllable optical output field in $\mathcal{PT}$-symmetric-like optomechanical system}

\begin{figure*}[ptb]
\includegraphics[bb=30 200 520 622,  width=0.4\textwidth, clip]{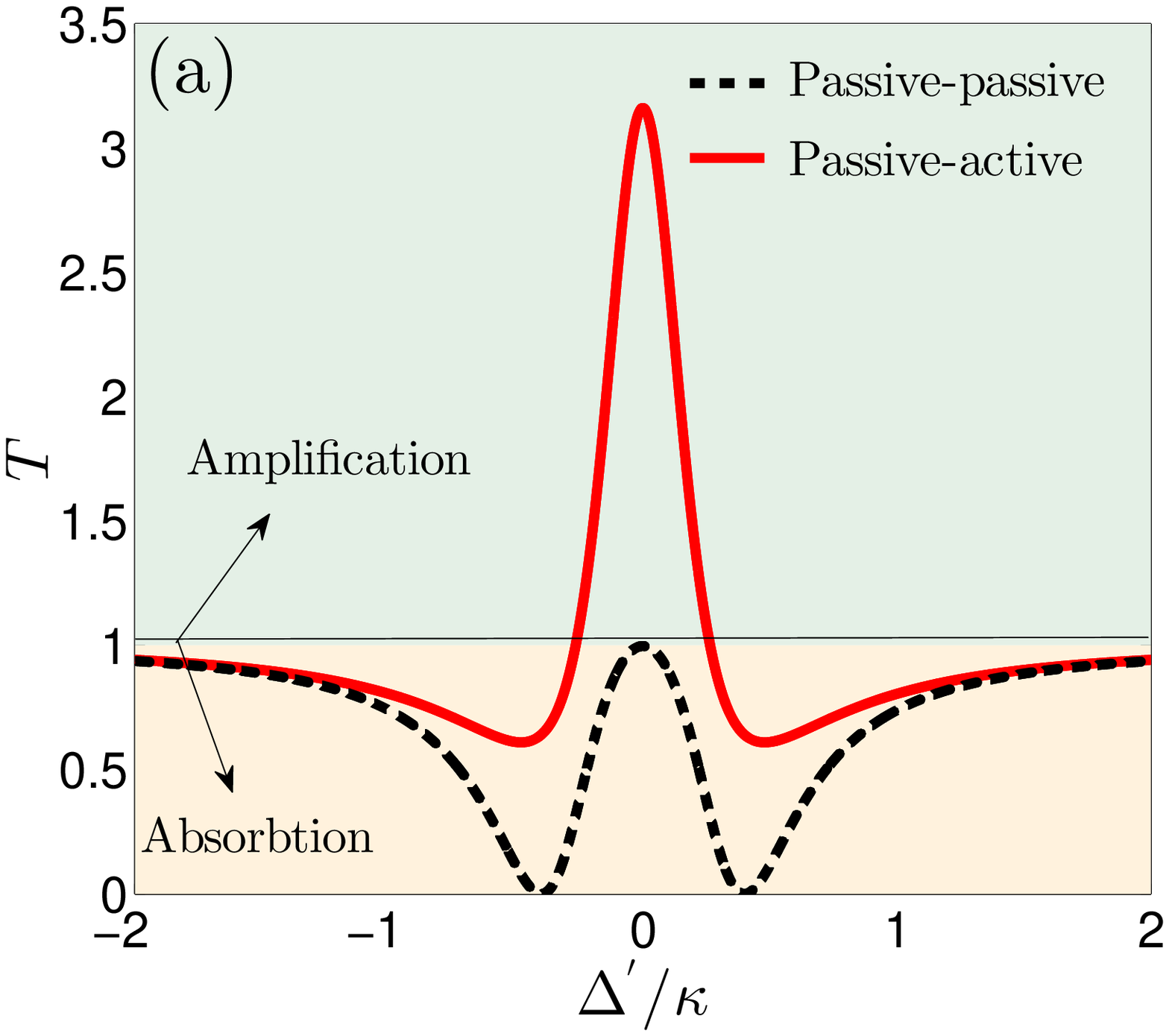}\includegraphics[bb=30 200 520 622,  width=0.4\textwidth, clip]{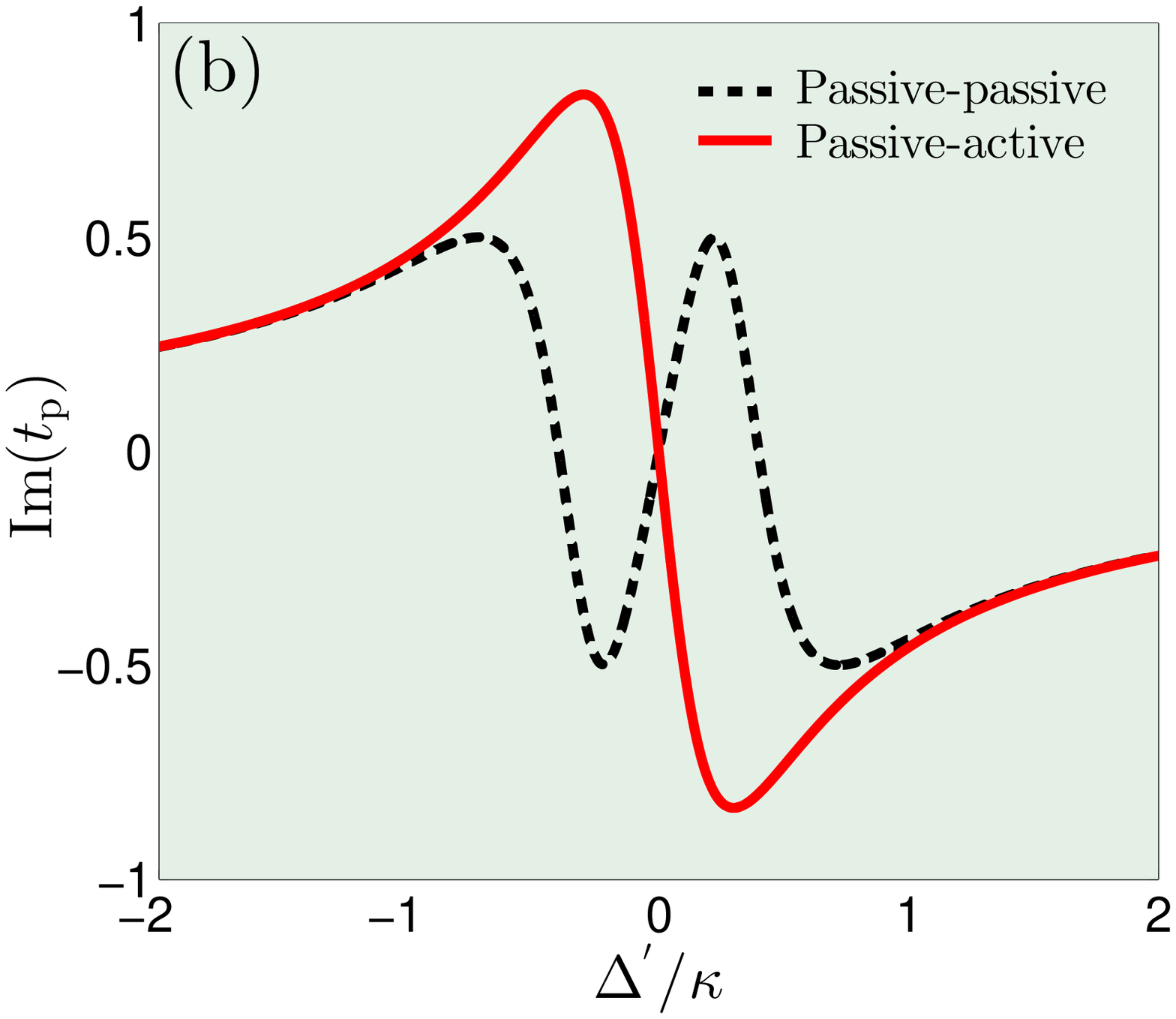}
\includegraphics[bb=30 200 520 622,  width=0.4\textwidth, clip]{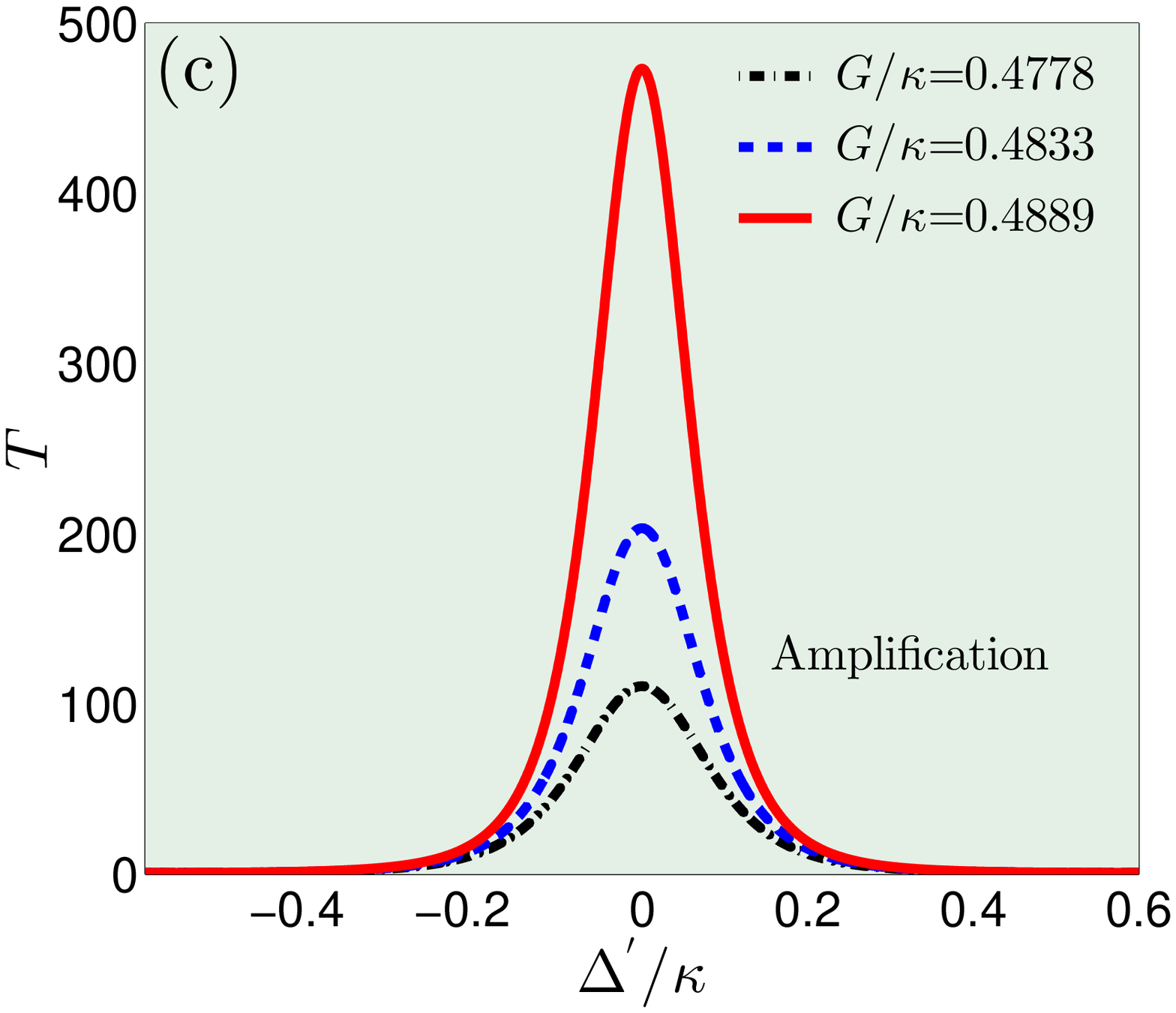}\includegraphics[bb=30 200 520 622,  width=0.4\textwidth, clip]{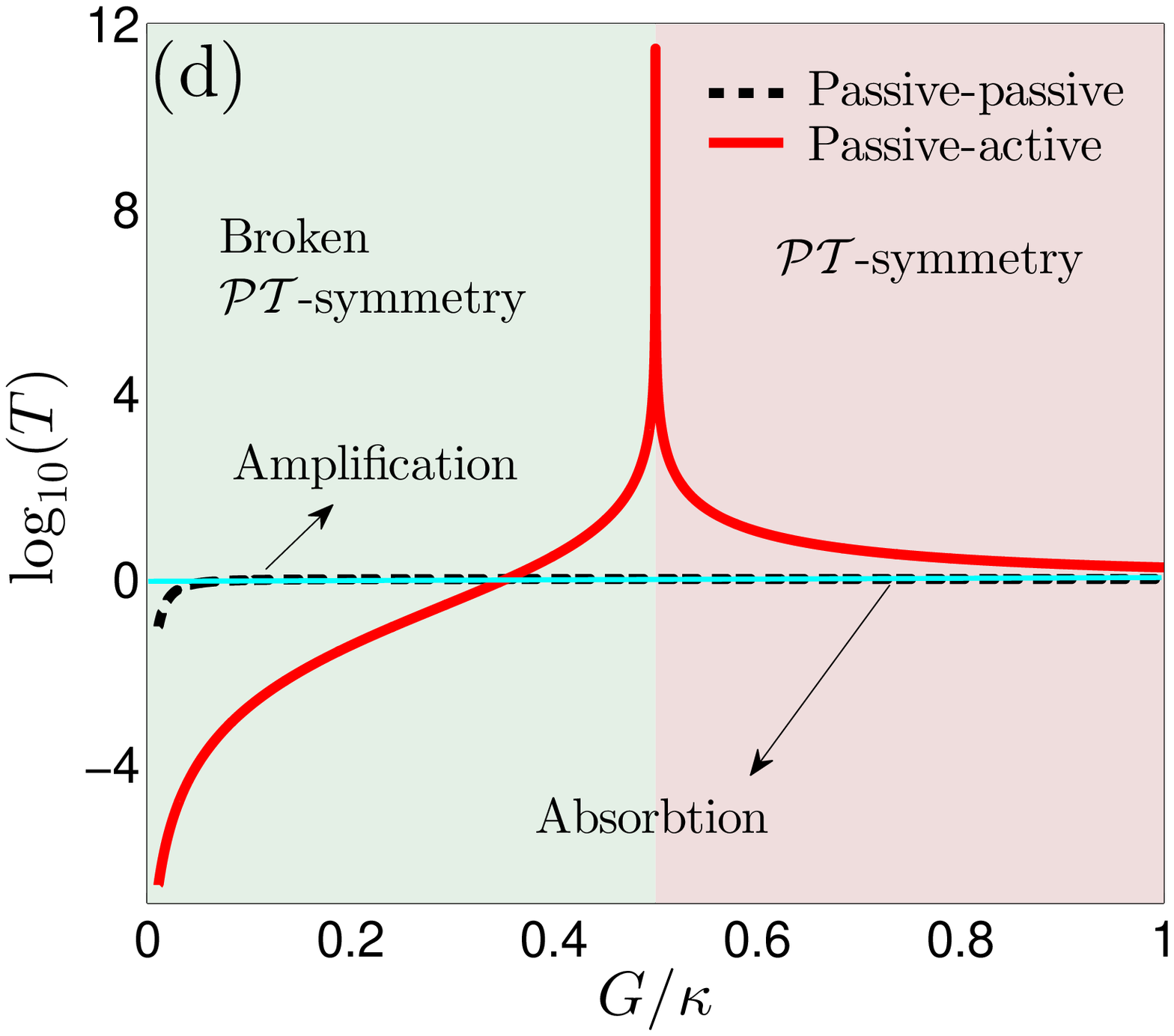}\caption{(Color online) The transmission coefficient $T$ and the dispersion Im($t_{\mathrm{p}}$) versus detuning $\Delta^{^{\prime}}$ with $G/\kappa=0.4$ are plotted in (a) and (b), respectively. (c) Optical amplification with different coupling strengths $G$ ($G/\kappa=0.4778,0.4833$,$\ $and $0.4889$). (d) The logarithm of the transmission coefficient $T$ with $\Delta^{^{\prime}}=0$. The depicted results are for the balanced gain and loss condition (i.e, $\kappa=\gamma$). The other system parameters are $\gamma_{\mathrm{m}}=1$ MHz, $\kappa/\gamma_{\mathrm{m}}$=900, and $\eta=0.5$. We use word ``Passive-passive" (``Passive-active") to represent conventional optomechanical systems ($\mathcal{PT}$-symmetric-like optomechanical system), respectively. Parameter $\gamma_{\mathrm{m}}$ is the mechanical dissipation rate for conventional optomechanical systems.}%
\label{fig6}%
\end{figure*}

According to the input-output theory~\cite{input-output}, the output optical field from the optomechanical cavity is given by%
\begin{equation}
\left\langle a_{\mathrm{out}}\right\rangle +\varepsilon_{\mathrm{c}%
}+\varepsilon_{\mathrm{p}}e^{-i\delta t}=\kappa_{\mathrm{e}}\left\langle
a\right\rangle , \label{input}%
\end{equation}
from which the transmission coefficient at the frequency of the probe field ($\omega_{\mathrm{p}}$) can be derived as (normalized to the input power of optical probe field)%
\begin{align}
t_{\mathrm{p}}  &  =\frac{\kappa_{\mathrm{e}}\left\langle A\right\rangle
-\varepsilon_{\mathrm{p}}}{\varepsilon_{\mathrm{p}}}=\frac{\kappa_{\mathrm{e}%
}\left\langle A\right\rangle }{\varepsilon_{\mathrm{p}}}-1\nonumber\\
&  =\frac{\eta\kappa\left(  i\omega_{2}-\frac{\gamma}{2}\right)  }{\left(
i\omega_{1}+\frac{\kappa}{2}\right)  \left(  i\omega_{2}-\frac{\gamma}%
{2}\right)  +\left\vert G\right\vert ^{2}}-1,
\end{align}
and the corresponding power transmission coefficient is given by%
\begin{equation}
T=\left\vert \frac{\eta\kappa\left(  i\omega_{2}-\frac{\gamma}{2}\right)
}{\left(  i\omega_{1}+\frac{\kappa}{2}\right)  \left(  i\omega_{2}%
-\frac{\gamma}{2}\right)  +\left\vert G\right\vert ^{2}}-1\right\vert ^{2}.
\label{tran}%
\end{equation}
Without loss of generality, in the following discussions the effective detuning $\Delta$ between the cavity field and the control field is fixed at the frequency of the mechanical resonator, i.e., $\Delta=\omega_{\mathrm{m}}$. So, $\omega_{1}=\omega_{2}=\omega_{\mathrm{m}}-\delta=-$ $\Delta^{^{\prime}}$. where, $\Delta^{^{\prime}}$ represents the frequency detuning between the probe field and the cavity field. Equation~(\ref{tran}) can then be further expressed as%
\begin{equation}
T=\left\vert \frac{\eta\kappa\left(  i\Delta^{^{\prime}}+\frac{\gamma}%
{2}\right)  }{\left(  i\Delta^{^{\prime}}-\frac{\kappa}{2}\right)  \left(
i\Delta^{^{\prime}}+\frac{\gamma}{2}\right)  +\left\vert G\right\vert ^{2}%
}+1\right\vert ^{2}. \label{Transimission}%
\end{equation}
By using Equation~(\ref{Transimission}), in the following we will discuss how to achieve controllable optical amplification, absorption, and group delay in the $\mathcal{PT}$-symmetric-like optomechanical system.

\subsection{Balanced gain-to-loss case}

We first consider a system with balanced gain and loss. The transmission coefficient $T$ versus the detuning $\Delta^{^{\prime}}$ is calculated and shown in Fig.~\ref{fig6}(a) for conventional optomechanical systems (represented by a black-dashed curve) and the proposed $\mathcal{PT}$-symmetric-like optomechanical system (represented by a red-solid curve), respectively. Here, the optomechanical coupling strength $G=0.4\kappa$ is assumed in both systems.

For conventional optomechanical systems, an Autler-Townes-splitting-like spectrum is observed~\cite{stcoup1,stcoup2,stcoup3}. However, for our $\mathcal{PT}$-symmetric-like optomechanical system, a remarkable probe amplification can be established between the two Autler-Townes absorption dips, where the peak is located at $\Delta^{^{\prime}}=0$. The dispersion [determined by the imaginary part of $t_{\mathrm{p}}$, i.e., Im($t_{\mathrm{p}}$)] behavior for both cases are also studied and shown in Fig.~\ref{fig6}(b). Around the parameter regime $\Delta^{^{\prime}}=0$, the conventional optomechanical systems exhibit an anomalous dispersive behavior (represented by a black-dashed curve) while the $\mathcal{PT}$-symmetric-like optomechanical system exhibits a normal dispersion (represented by a red-solid curve).

To see how this amplification is affected by the coupling strength $G$, the transmission coefficient $T$ versus detuning $\Delta^{^{\prime}}$ for different coupling strengths $G$ (e.g. $G/\kappa=0.4778,G/\kappa=0.4833,G/\kappa=0.4889$) are calculated and shown in Fig.~\ref{fig6}(c). We can see that the amplification can be controlled by changing the coupling strength $G$. Also, the amplification is very sensitive to $G$ when the system is brought to the vicinity of the phase transition point $G=0.5\kappa$. As shown in Fig.~\ref{fig6}(d), the black-dashed curve represents conventional optomechanical system, in which the transmission coefficient monotonically increases to one without any signal amplification. However, for the $\mathcal{PT}$-symmetric-like optomechanical system (the red-solid curve), strong amplification is achieved near the phase transition point. Additionally, a perfect optical absorption can be achieved (e.g. $T=10^{-5}$) in the broken
$\mathcal{PT}$-symmetry regime.

Note that a $\mathcal{PT}$-symmetric-like optomechanical system can be used to realize strong optical amplification with a red-detuned control field. According to Eq.~(\ref{Transimission}), the transmission coefficient $T$ at $\Delta^{^{\prime}}=0$ can be given by
\begin{equation}
T=\left\vert \frac{(2\eta-1)\frac{\kappa\gamma}{4}+\left\vert G\right\vert
^{2}}{\left\vert G\right\vert ^{2}-\frac{\kappa\gamma}{4}}\right\vert ^{2}.
\label{AmplificationAA}%
\end{equation}
which indicates a strong amplification near the phase transition point $G=\sqrt{\kappa\gamma}/2$. The huge amplification at this point is clearly seen in Fig.~\ref{fig6}(d). Figure~\ref{fig6} clearly shows that both tunable optical signal amplification and tunable absorption can be observed in the $\mathcal{PT}$-symmetric-like optomechanical system. The transition from optical absorption (amplification) to amplification (absorption) is controlled by increasing (decreasing) the optomechanical coupling strength $G$.
\begin{figure}[ptb]
\includegraphics[bb=0 120 715 450,  width=9cm, clip]{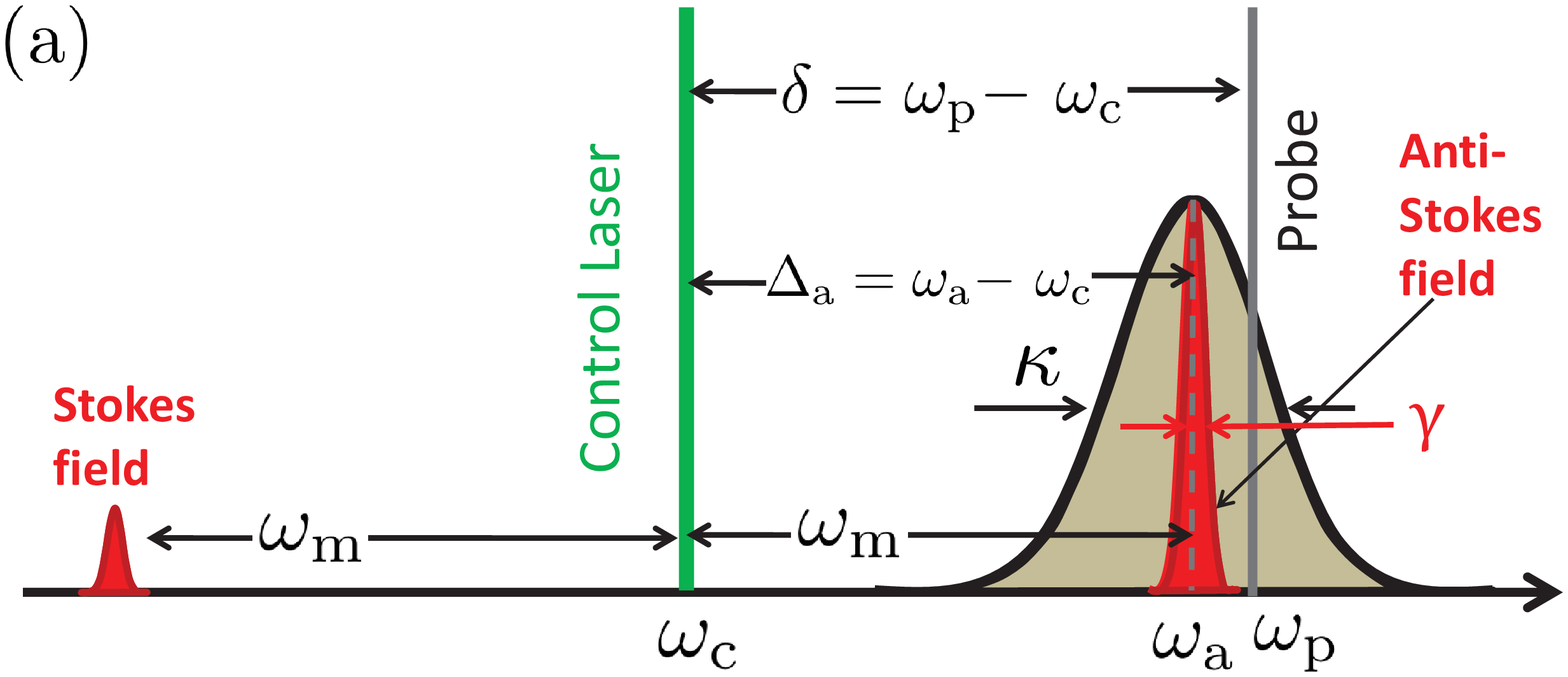}
\includegraphics[bb=0 105 694 450,  width=9cm, clip]{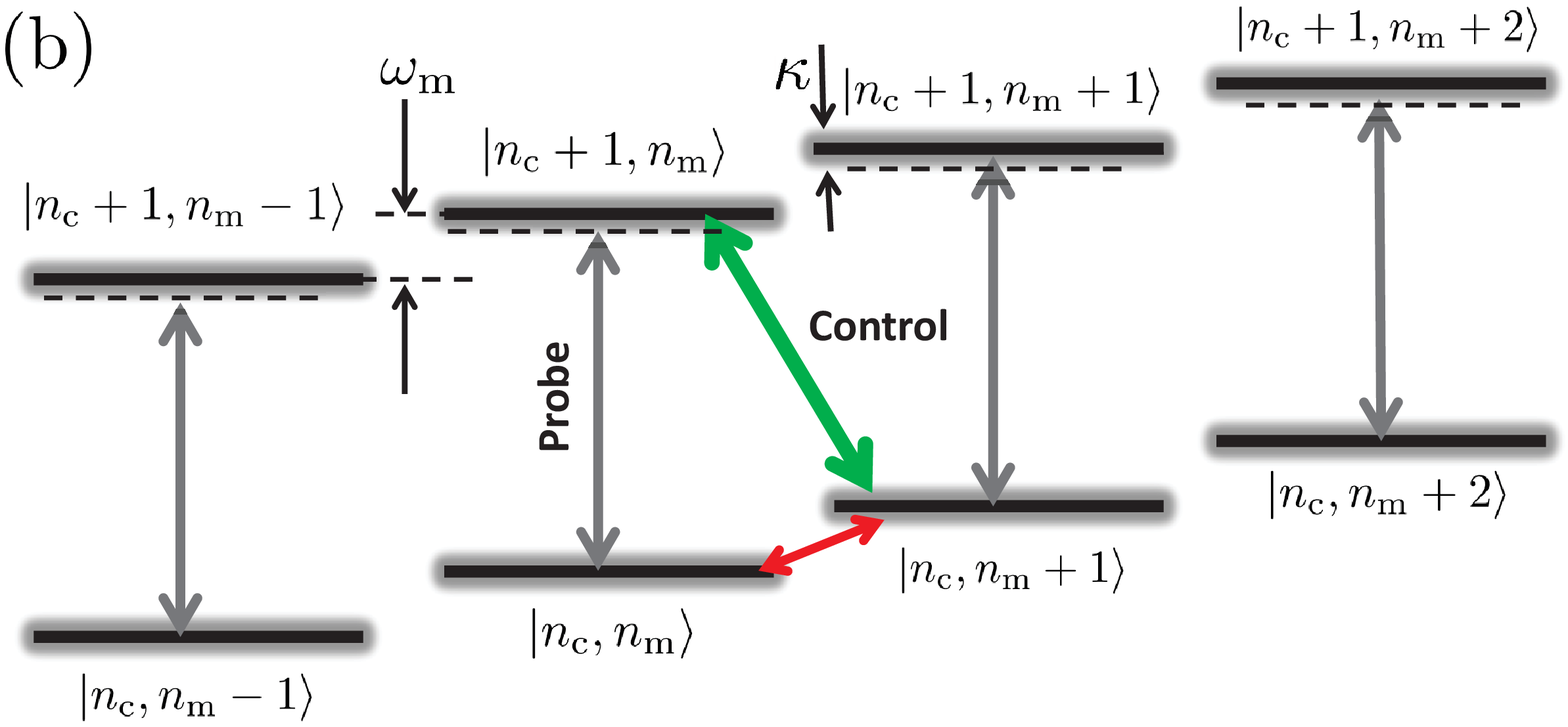}\caption{(Color online) (a) Schematic diagram explaining the various frequencies and the processes of pumping and probing. The system is driven by a control field at a frequency $\omega_{\mathrm{c}}$, which is red-detuned from the cavity resonant frequency $\omega_{\mathrm{a}}$ by about the mechanical resonant frequency $\omega_{\mathrm{m}}$. A weak probe field with frequency (i.e., $\omega_{\mathrm{p}}=$ $\omega_{\mathrm{c}}+\delta$) is used to probe the cavity response. (b) Energy levels diagram explain the enhanced raman scattering and stimulated-emission-like process in the $\mathcal{PT}$-symmetric-like optomechanical system. The product states $\left\vert n_{\mathrm{c}},n_{\mathrm{m}}\right\rangle $ are characterized by $n_{\mathrm{c}}$\ cavity photons and $n_{\mathrm{m}}$ mechanical phonons. Also, $\kappa$ ($\gamma$) represents the loss (gain) of the cavity (mechanical) mode, respectively.}%
\label{fig7}%
\end{figure}

The physical mechanism of photon amplification in $\mathcal{PT}$-symmetric-like optomechanical system is illustrated by the driving-strategy and simplified energy-level transition pictures in Fig.~\ref{fig7}(a) and Fig.~\ref{fig7}(b), respectively. As shown in Fig.~\ref{fig7}(a), the system is driven by a strong control field at frequency $\omega_{\mathrm{c}}$, which is red-detuned from the cavity resonant frequency $\omega_{\mathrm{a}}$ by about the mechanical resonant frequency $\omega_{\mathrm{m}}$. A weak probe field at the frequency $\omega_{\mathrm{p}}=$ $\omega_{\mathrm{c}}+\delta$ probes the modified cavity resonance. The strong control field exerts a radiation pressure on the oscillating mirror that induces a controllable coupling between the cavity and mechanical modes. Simultaneously, the vibrating mechanical mode photoelastically generates an optical grating that is capable of scattering the control optical field~\cite{bahl,chunhua}. Therefore, the Stokes and anti-Stokes fields [red-shaded areas in Fig.~\ref{fig7}(a)] build up at $\omega_{\mathrm{c}}\pm\omega_{\mathrm{m}}$ around the strong-driving field. Since the system works in the resolved-sideband regime, $\omega_{\mathrm{m}}>(\kappa,\gamma)$, the Stokes field at $\omega_{\mathrm{c}}-\omega_{\mathrm{m}}$ is strongly suppressed because it is off-resonant with the cavity, whereas the anti-Stokes field at $\omega_{\mathrm{c}}+\omega_{\mathrm{m}}$ is enhanced.

Note that the $\mathcal{PT}$-symmetric-like optomechanical system proposed in this work differs from conventional optomechanical systems by the controllably-introduced mechanical gain. The generation of such mechanical gain can be viewed as a continuous coherent phonon bath pumping the mechanical mode, which in turn greatly enhances the photoelastic scattering and the anti-Stokes field. Since the probe field is resonant to the anti-Stokes scattered optical field, it allows for stimulated-emission-like photon amplification of the probe field, which is
analogous to the stimulated emission in atoms. Here, the probe response is controlled by the control field and the mechanical gain. The resulting stimulated emission of the intra-cavity field forms an amplification window when the two-photon resonance condition is met.

\begin{figure*}[ptb]
\includegraphics[bb=40 199 547 620,  width=0.4\textwidth, clip]{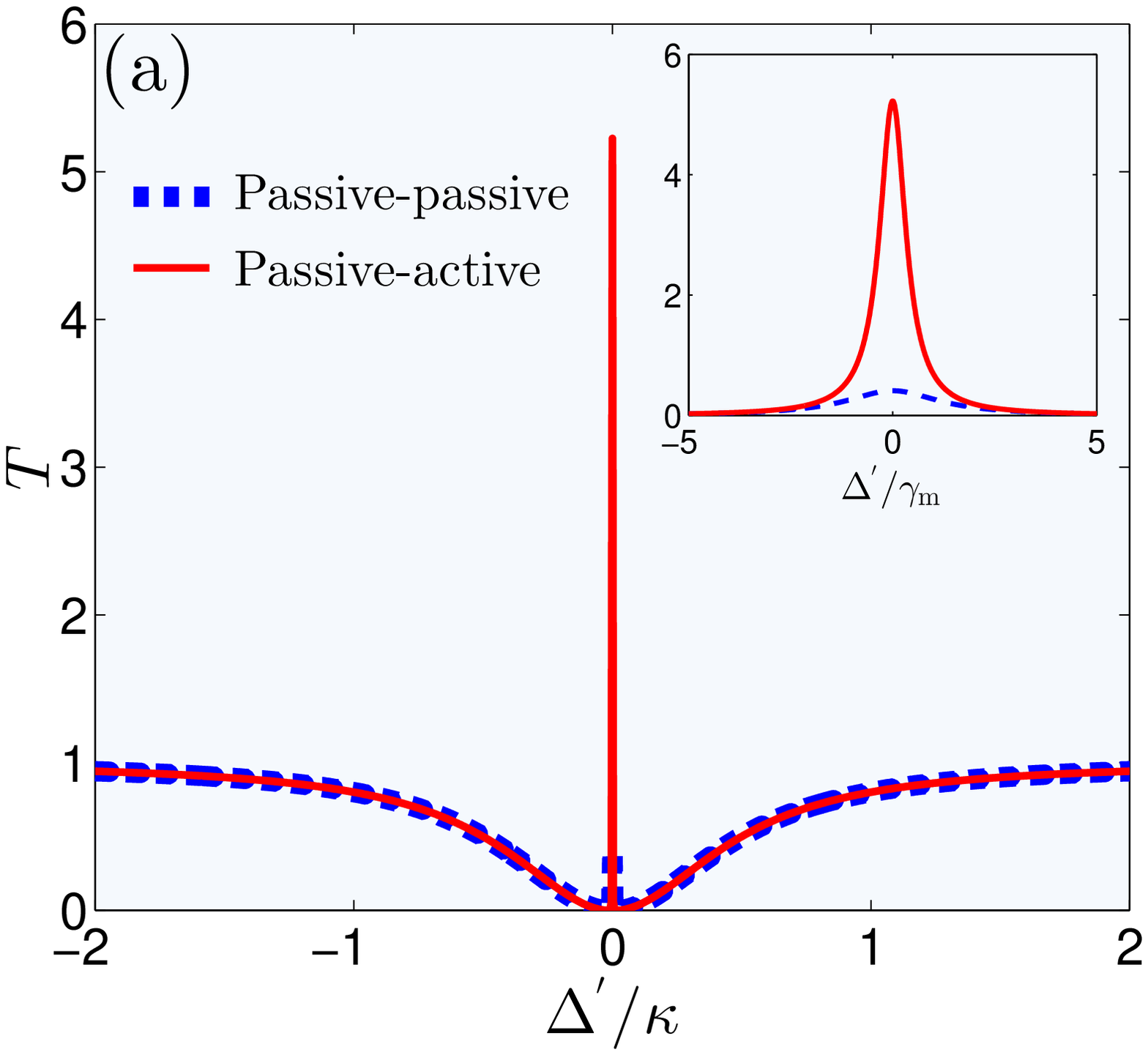}\includegraphics[bb=40 199 547 620,  width=0.4\textwidth, clip]{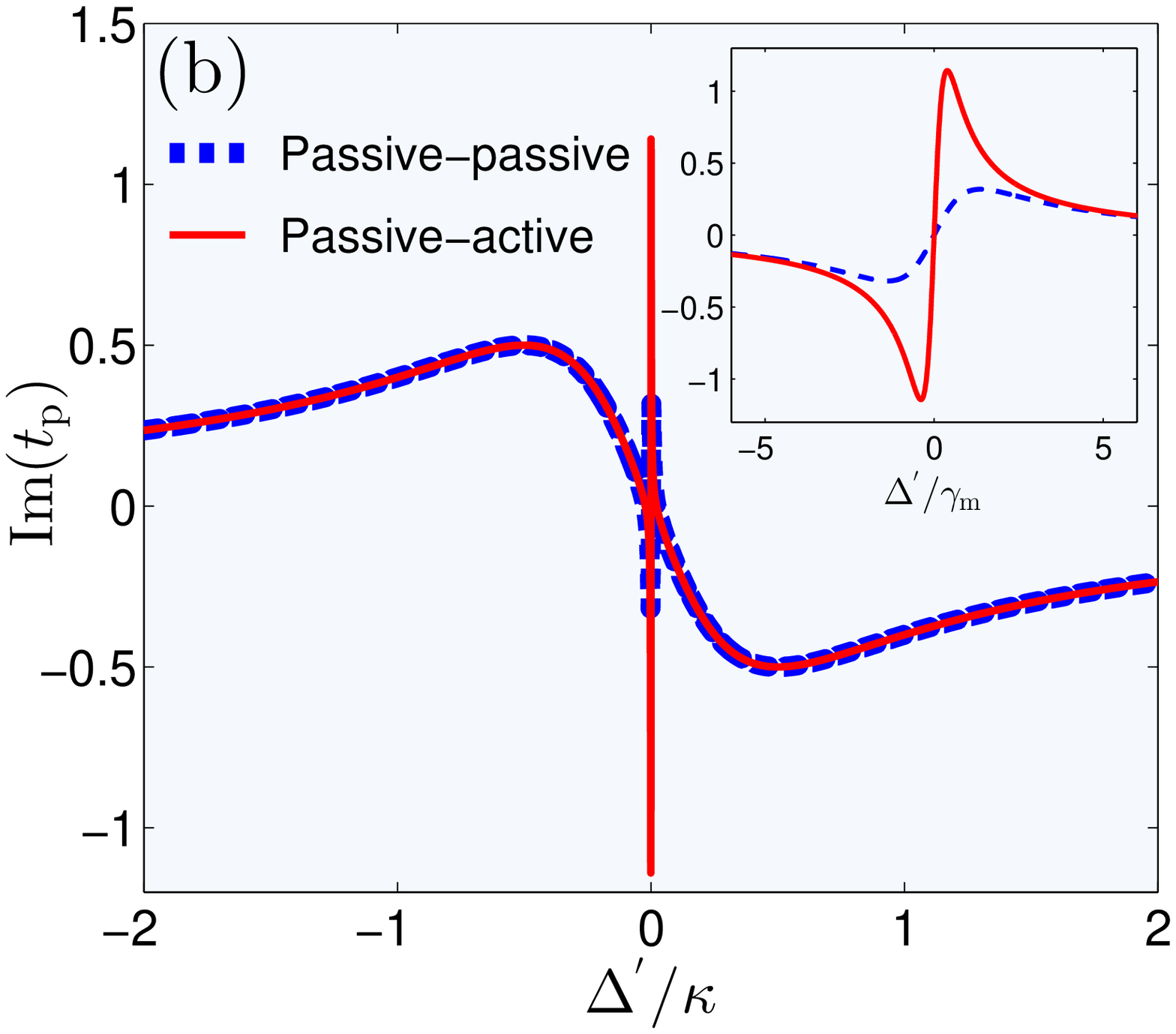}\caption{(Color online) The transmission coefficient $T$ and the dispersion Im($t_{\mathrm{p}}$) versus detuning $\Delta^{^{\prime}}$ are plotted in (a) and (b), respectively. The blue-dashed curve (red-solid curve) represents conventional optomechanical system ($\mathcal{PT}$-symmetric-like optomechanical system). The other system parameters are $\gamma_{\mathrm{m}}=1$ MHz, $G/\gamma_{\mathrm{m}}=20$, $\kappa/\gamma_{\mathrm{m}}$=900, $\gamma/\gamma_{\mathrm{m}}=1$, and $\eta=0.5$. We use word ``Passive-passive" (``Passive-active") to represent conventional optomechanical systems ($\mathcal{PT}$-symmetric-like optomechanical system), respectively. Parameter $\gamma_{\mathrm{m}}$ is the mechanical dissipation rate for conventional optomechanical systems.}%
\label{fig8}%
\end{figure*}

One can also understand the effect of photon amplification through the energy-level transition picture. As shown in Fig.~\ref{fig7}(b), the
optomechanical states consist of product states $\left\vert n_{\mathrm{c}},n_{\mathrm{m}}\right\rangle $, where $n_{\mathrm{c}}$\ and $n_{\mathrm{m}}$ are the numbers of cavity photons and mechanical phonons, respectively. Here, a pure cavity mode excitation is represented by a black vertical arrow increasing only the number of photons in the cavity. In contrast, a control field tuned at $\omega_{\mathrm{c}}\simeq\omega_{\mathrm{a}}-\omega_{\mathrm{m}}$ (optimal detuning) creates (annihilates) one photon (phonon) number by one, which is represented by a green diagonal arrow. The photons of the control field are up-converted and scattered into the anti-Stokes line at $\omega_{\mathrm{c}}+\omega_{\mathrm{m}}$, approximately matching the cavity resonant frequency $\omega_{\mathrm{a}}$. The mechanical gain and the coherent phonon pump are shown by the red diagonal arrow. The mechanical gain supports enough phonons for the consumption during the anti-Stokes scattering process. Thus, the photoelastic scattering and the anti-Stokes field are greatly enhanced. Because the probe field is coherently resonant with the anti-Stokes field, a stimulated-emission-like process is observed, as well as strong amplification of the probe field.

From the above discussion we see that the physical mechanism of mechanical-gain-induced strong optical amplification is totally different from the Raman process~\cite{raman} or four-wave mixing~\cite{fourw}, which rely on the optical nonlinear response of the medium. The strong third-order nonlinear susceptibility is also not needed here. Our method is suitable for any spectral band and for optical signals with different wavelength. It can also exhibit a broadband optical amplification~\cite{mo}. The great amplification discussed here is enabled by the enhanced photoelastic-scattering and anti-Stokes field around the phase transition point.

\subsection{Unbalanced gain-to-loss case}

Optomechanically-induced-transparency is useful for slowing and switching probe lights, and it may be further developed for on-chip optical pulse storage. The underlying physics of OMIT is similar to electromagnetically-induced transparency (EIT) in atomic physics. Group delay is the most prominent feature of EIT. As is well known in conventional optomechanical systems, the group delay time $\tau$ is mainly determined by the mechanical dissipation rate $\gamma_{\mathrm{m}}$ ($\tau\sim1/\gamma_{\mathrm{m}}$). Despite significant advances in the performance of optomechanical devices in the past few years, the realization of both high optical transmission and long group delay still remain a challenge~\cite{peit1,peit3,peit5}. Due to the resolved-sideband condition, a high mechanical frequency is required, which inevitably leads to a high mechanical dissipation rate, and thereby small group delay. Even though the optical transmission can be increased by a larger optomechanical cooperativity, the broadening of the transparency window indicates further decrease in group delay. In the following, we will discuss an OMIT-like spectrum in the $\mathcal{PT}$-symmetric-like optomechanical system. In contrast to conventional optomechanical systems, a small gain for the mechanical mode is introduced ($\left\vert \gamma\right\vert=\gamma_{\mathrm{m}}$), so that the system works in the deeply-unbalanced gain and loss regime. We will show that even a small amount of mechanical gain may greatly enhance the transmission center peak, which corresponds to a much sharper change in the dispersion curve and hence to an ultra-long group delay.

\begin{figure*}[ptb]
\includegraphics[bb=32 200 550 630,  width=0.4\textwidth, clip]{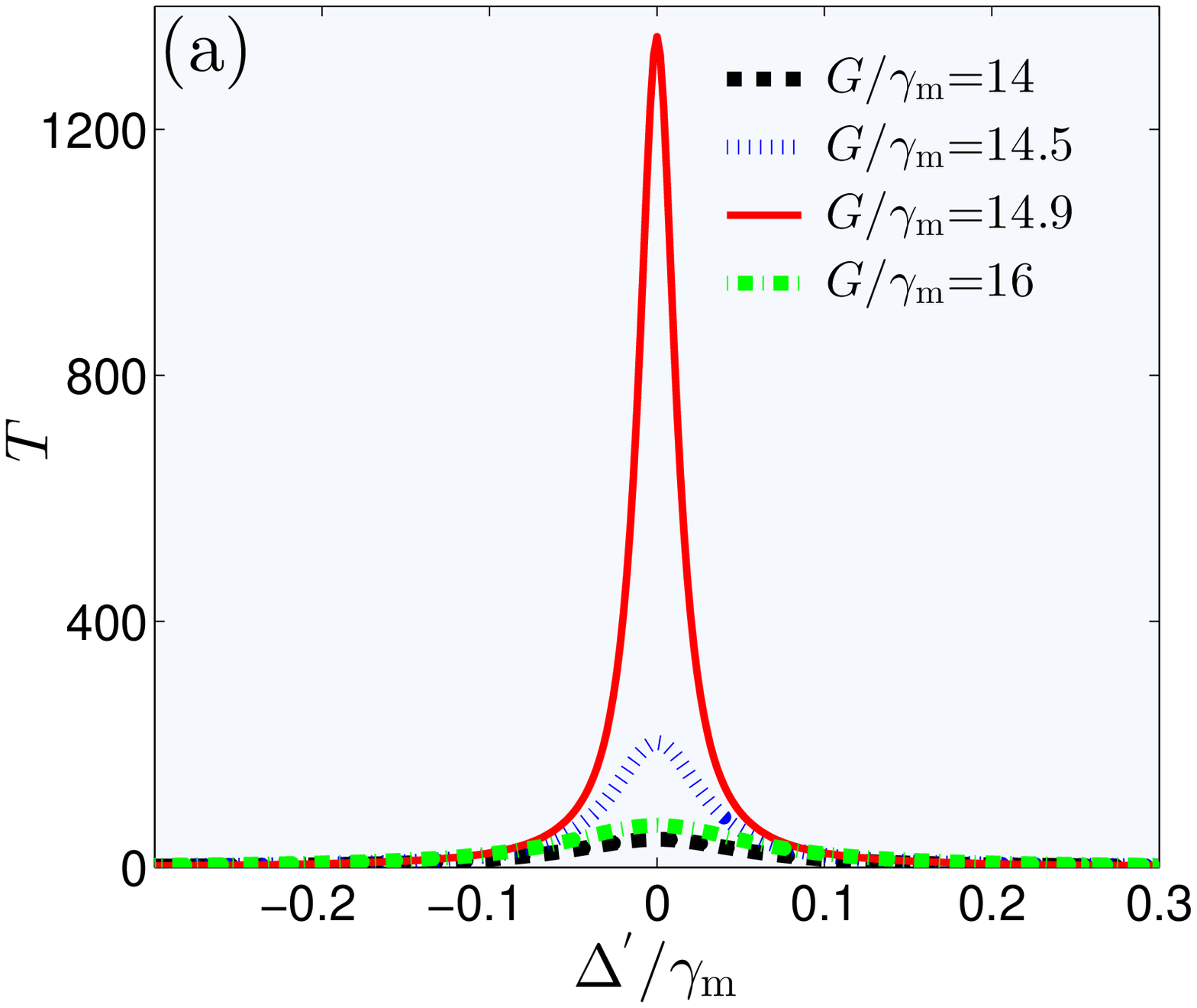}\includegraphics[bb=30 200 550 630,  width=0.4\textwidth, clip]{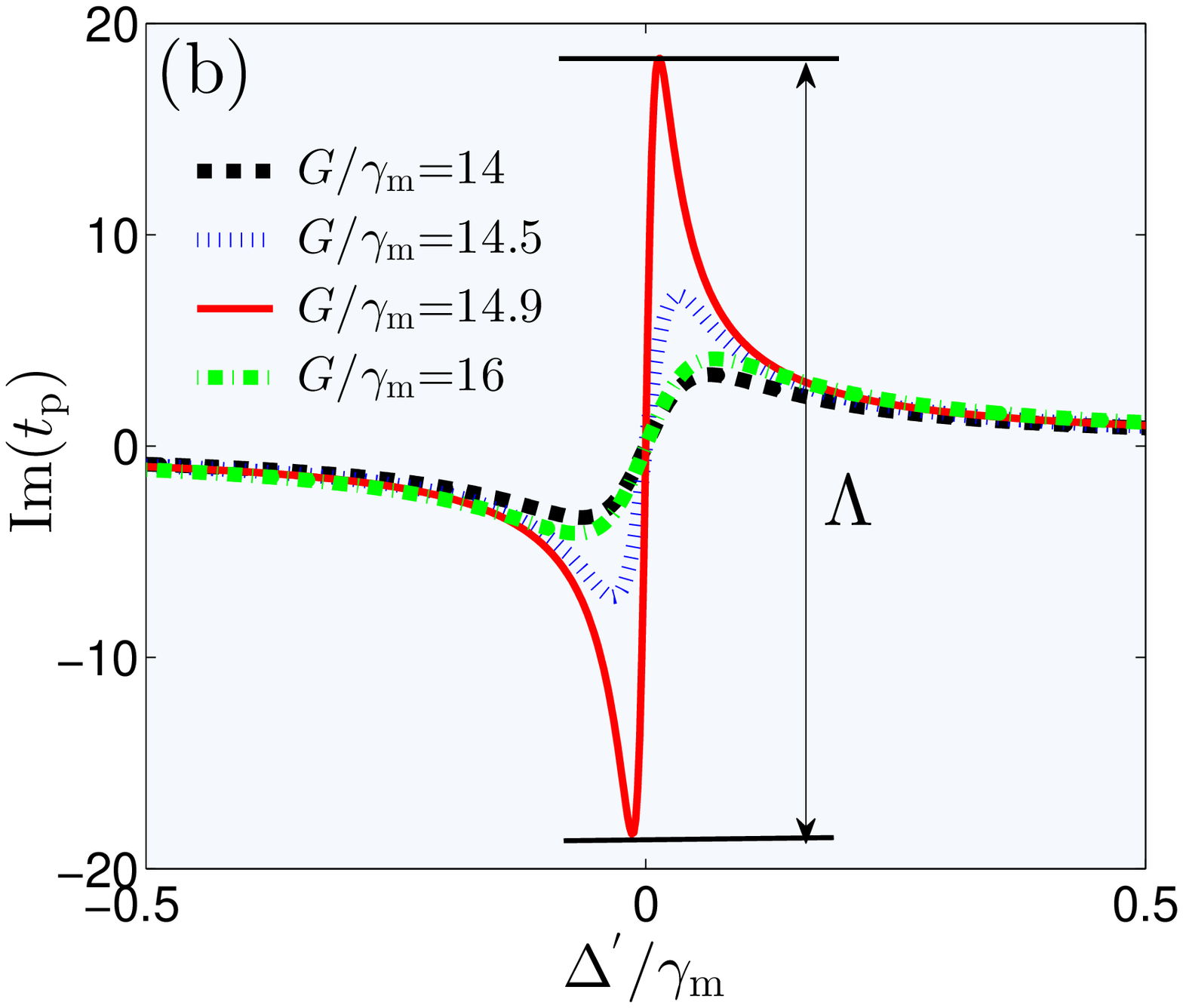}
\includegraphics[bb=30 200 550 630,  width=0.4\textwidth, clip]{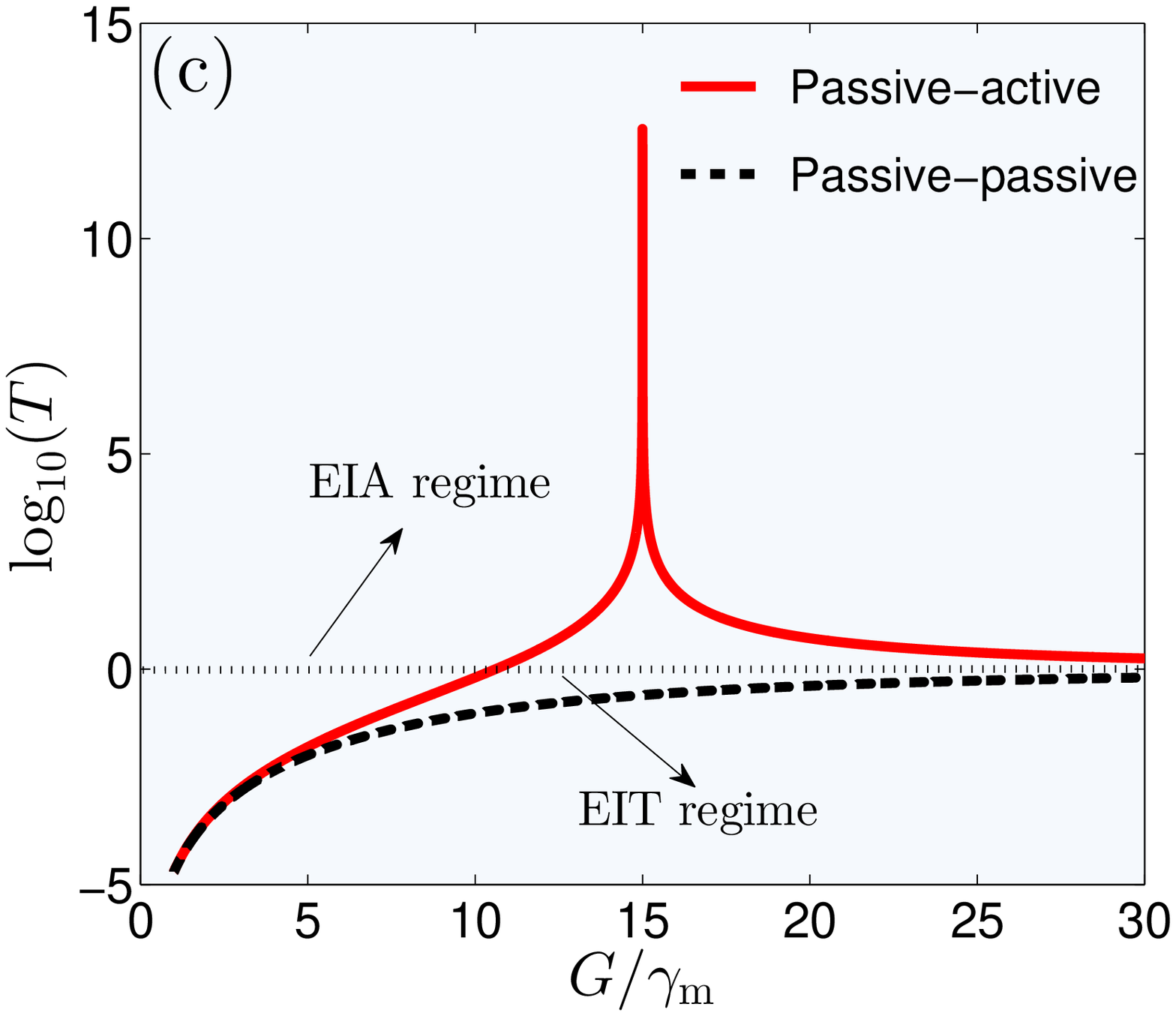}\includegraphics[bb=30 200 550 630,  width=0.4\textwidth, clip]{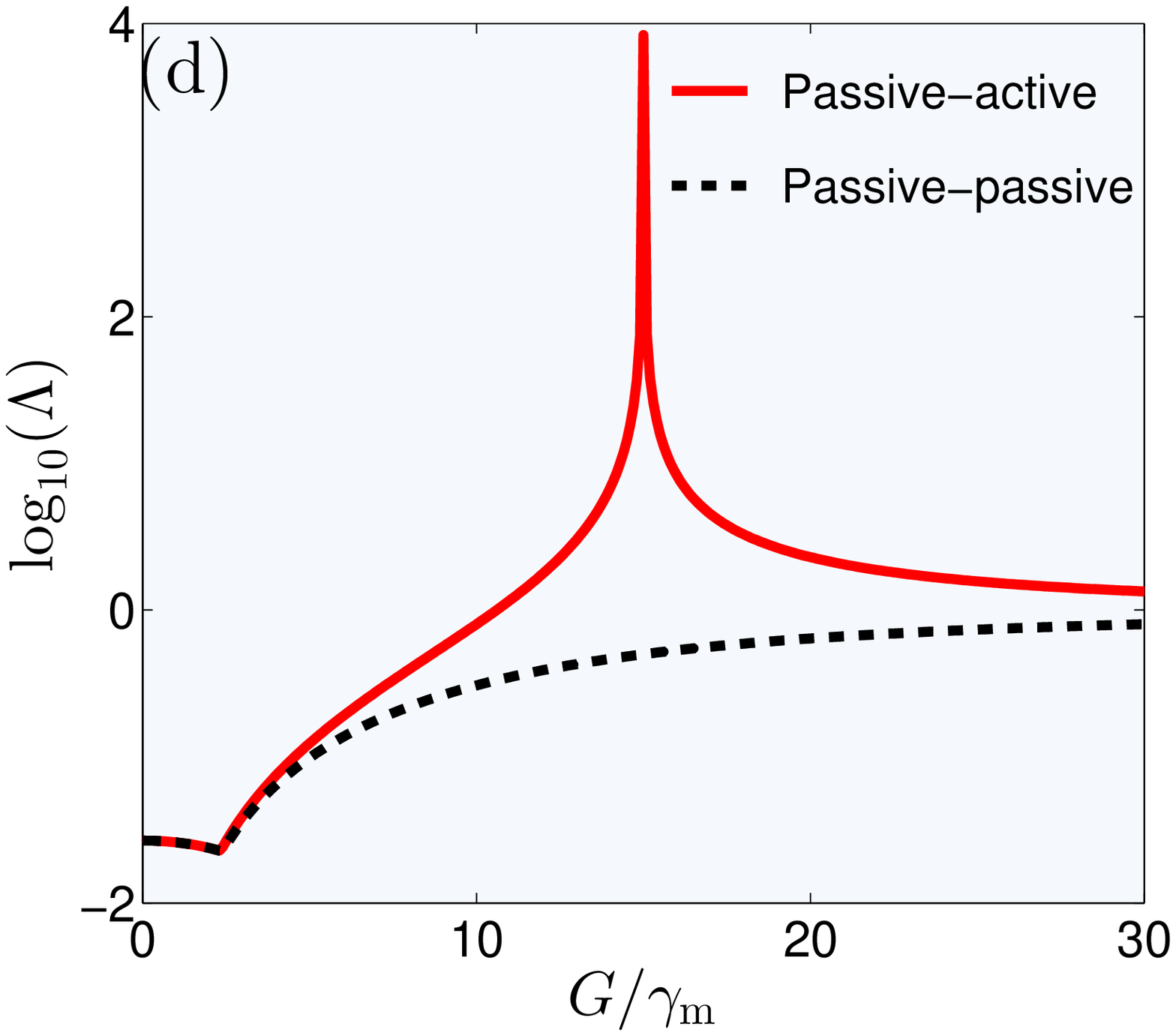}\caption{(Color online) The transmission coefficient $T$ and dispersion Im($t_{\mathrm{p}}$) versus the detuning $\Delta^{^{\prime}}$ with different coupling strengths $G/\gamma_{\mathrm{m}}=$($14,14.5,14.9$ and $16$) are plotted in (a) and (b), respectively. The logarithm of the transmission coefficient $T$ and the dispersion variation $\Lambda$ versus $G$ with $\Delta^{^{\prime}}=0$ are plotted in (c) and (d). The electromagnetically-induced-transparency (EIT) regime (corresponding to $T<1$) and the electromagnetically-induced-amplification (EIA) regime (corresponding to $T>1$) are labeled below and on the dash-dotted line (corresponding to $T=1$) in subfigure~(c), respectively. The other system parameters are $\gamma_{\mathrm{m}}=1$ MHz, $\kappa/\gamma_{\mathrm{m}}$=900, $\gamma/\gamma_{\mathrm{m}}=1$, and $\eta=0.5$. We use word ``Passive-passive" (``Passive-active") to represent conventional optomechanical systems ($\mathcal{PT}$-symmetric-like optomechanical system), respectively. Parameter $\gamma_{\mathrm{m}}$ is the mechanical dissipation rate for conventional optomechanical systems.}%
\label{fig9}%
\end{figure*}

\subsubsection{Optomechanically-induced-transparency-like spectra}

The transmission coefficient $T$ versus detuning $\Delta^{^{\prime}}$ is calculated and shown in Fig.~\ref{fig8}(a), where the optomechanical coupling strength is taken as $G=0.22\kappa$. For conventional optomechanical systems (the blue-dashed curve), one can obtain an EIT-like spectrum~\cite{peit1,peit3,peit4,peit5,peit6}. However, for our $\mathcal{PT}$-symmetric-like optomechanical system (the red-solid curve), a remarkable probe amplification is observed near $\Delta^{^{\prime}}=0$. The center peak is calculated and shown in the inset of Fig.~\ref{fig8}(a). The dispersion curves for conventional optomechanical systems and the $\mathcal{PT}$-symmetric-like optomechanical system are shown in Fig.~\ref{fig8}(b), where we see that both exhibit anomalous dispersion. The enhancement of the optical transmission [see the red-solid curve in Fig.~\ref{fig8}(a)] leads to an abrupt variation of the dispersion [see the red-solid curve in Fig.~\ref{fig8}(b)] and therefore a large optical group delay of the outgoing optical field. The center dispersion variation is shown in the inset of Fig.~\ref{fig8}(b).

Next, we discuss how the coupling strength $G$ affects the enhancement of the group delay of the optical field. The transmission coefficient $T$ versus detuning $\Delta^{^{\prime}}$ for different coupling strengths $G$ (e.g., $G/\gamma_{\mathrm{m}}=14,G/\gamma_{\mathrm{m}}=14.5,G/\gamma_{\mathrm{m}}=14.9,G/\gamma_{\mathrm{m}}=16$) are shown in Fig.~\ref{fig9}(a), which clearly show that the height of the center peak is sensitive to the coupling strength $G$. Combined with the dispersion curves shown in Fig.~\ref{fig9}(b), we see that both the center peaks of the transmission and variation of the dispersion can be greatly enhanced with a properly chosen $G$. Also, the enhancement of the transmission [see Fig.~\ref{fig9}(a)] and fast variation of the dispersion [see Fig.~\ref{fig9}(b)] are not linearly dependent on the coupling strength $G$; that is, there is a critical value of $G$ beyond which leads to a decrease in the enhancement.

In contrast with conventional optomechanical systems, the $\mathcal{PT}$-symmetric-like optomechanical system possesses one critical point
($G_{\mathrm{cp}}=\sqrt{\kappa\gamma_{\mathrm{m}}}/2$) in the broken $\mathcal{PT}$-symmetry regime. When $G$ is set to bring the system to the vicinity of the critical point, an enhancement of the probe field and sharp variation of the dispersion can be achieved. Here, the critical point is located at $G_{\mathrm{cp}}/\gamma_{\mathrm{m}}=15$. As shown in Fig.~\ref{fig9}(b), the increment $\Lambda$ is introduced to quantify the abruptness of the dispersions, and therefore the length of the group delay. The group delay can be intuitively judged by the value of $\Lambda$. The larger the positive $\Lambda$ is, the longer the group delay is.
\begin{figure*}[ptb]
\includegraphics[bb=17 196 550 630,  width=0.4\textwidth, clip]{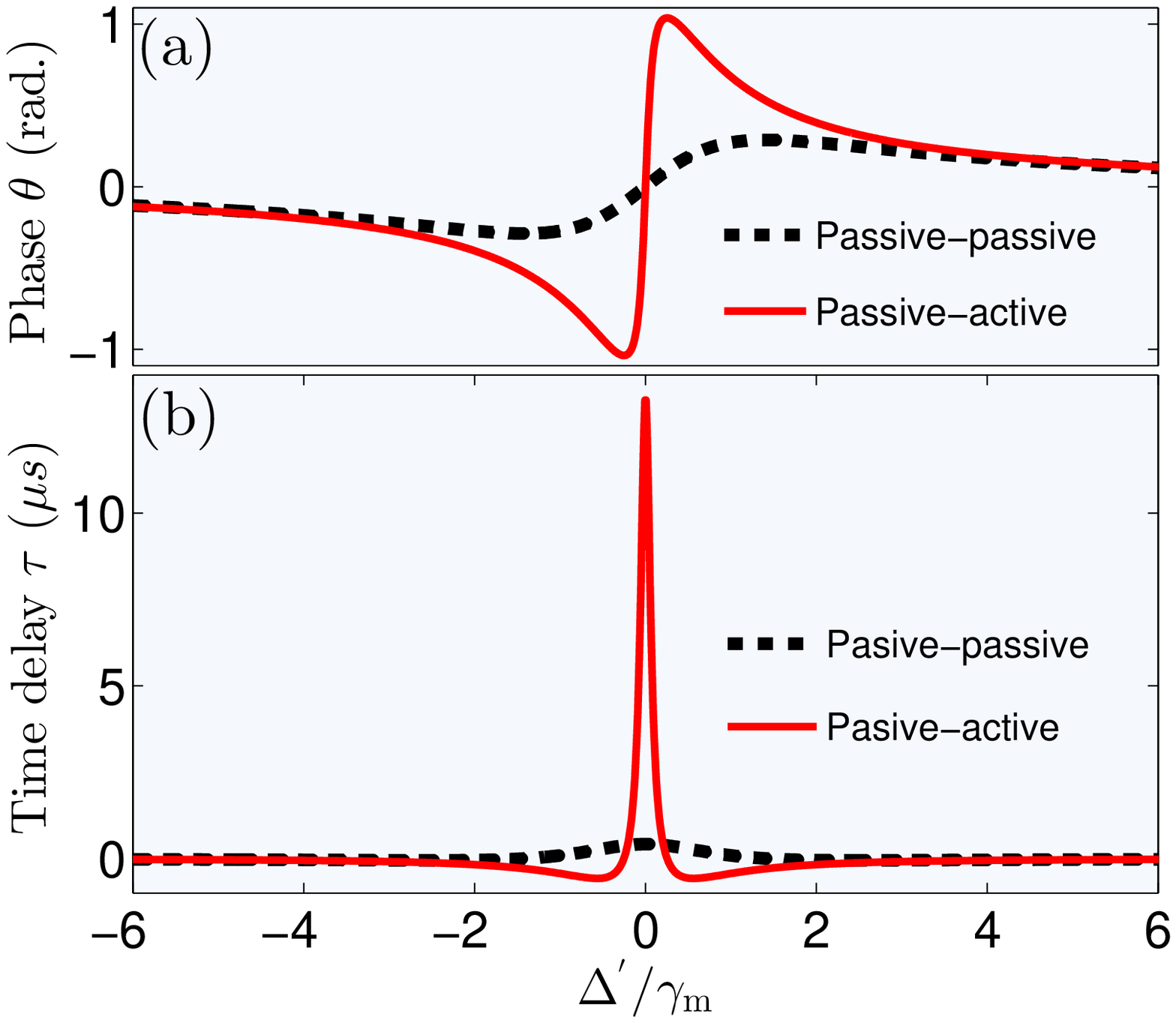}\includegraphics[bb=17 196 550 630,  width=0.4\textwidth, clip]{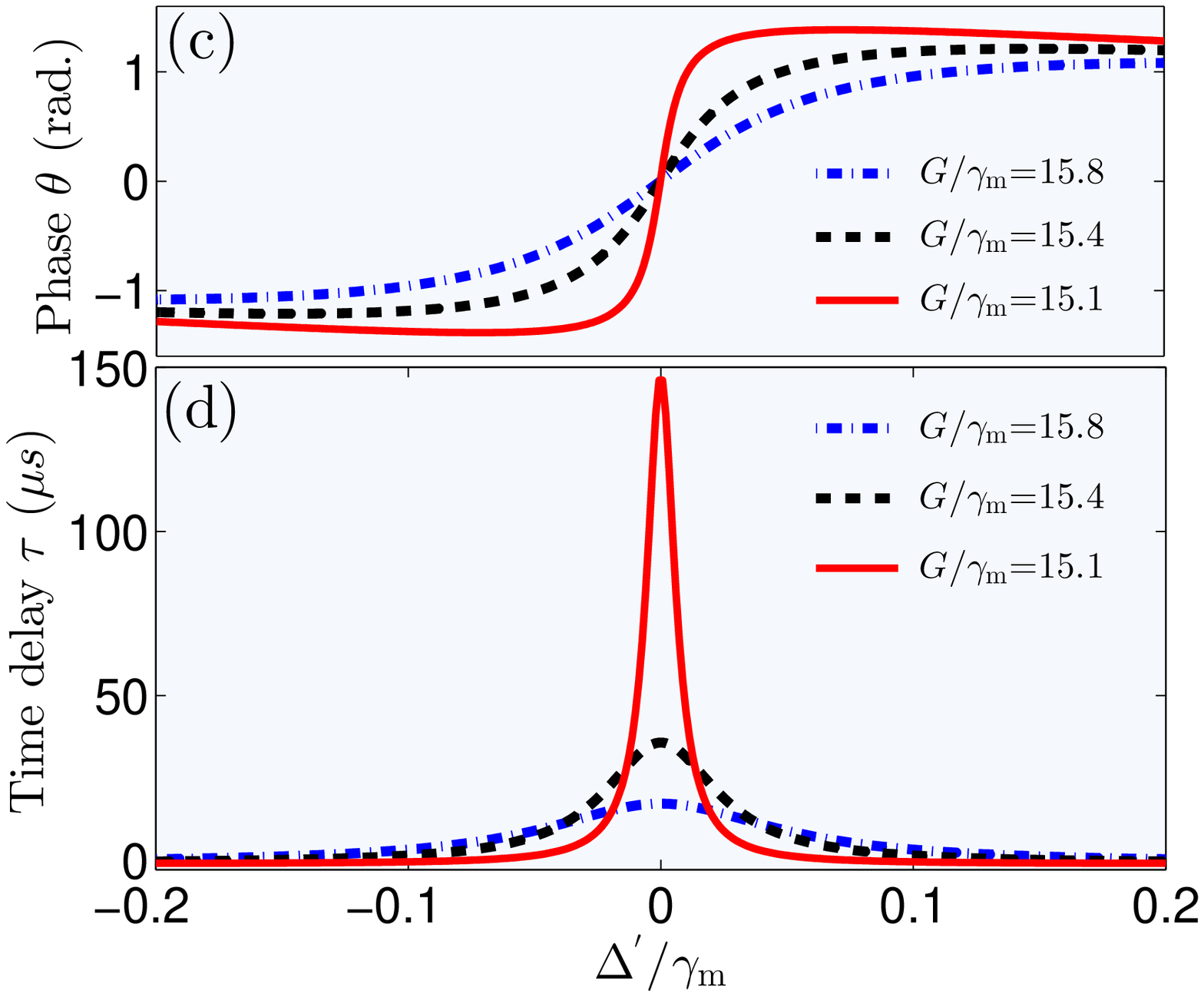}
\includegraphics[bb=17 196 550 630,  width=0.4\textwidth, clip]{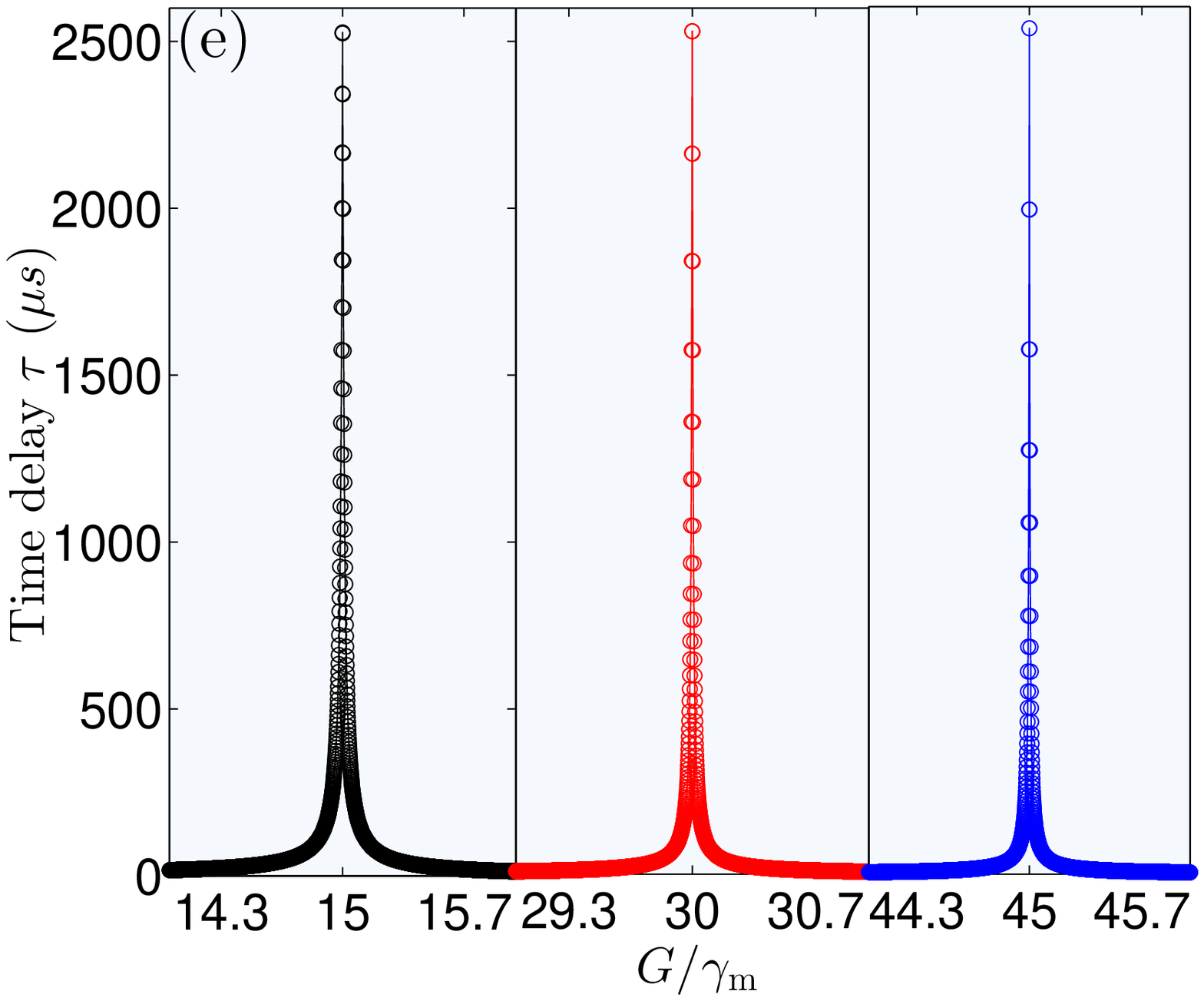}\includegraphics[bb=17 196 550 630,  width=0.4\textwidth, clip]{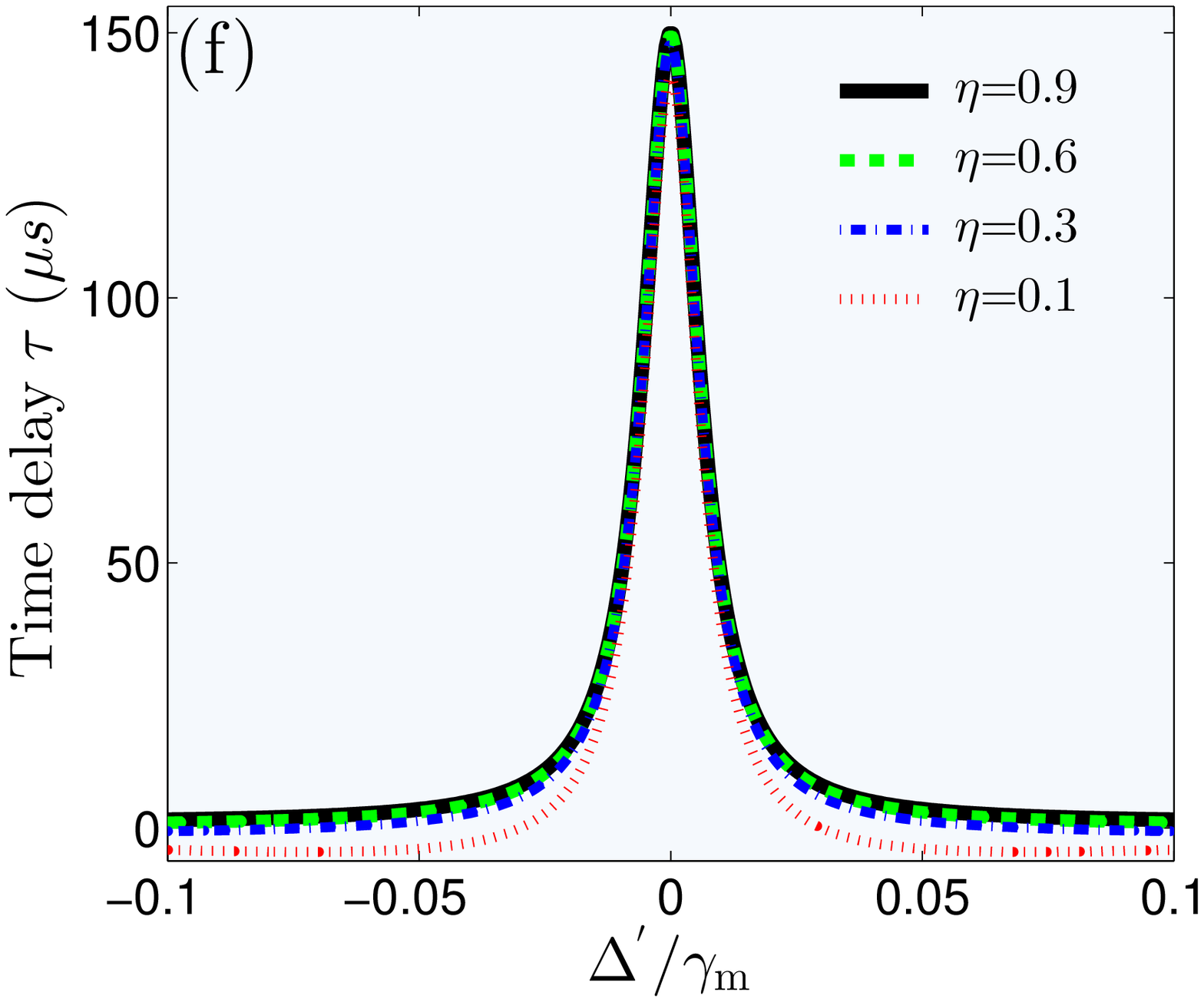}\caption{(Color online) Phase $\theta$ and group delay $\tau$ for conventional (see the black-dashed curve) and $\mathcal{PT}$-symmetric (see the red-solid curve) optomechanical systems versus $\Delta^{^{\prime}}$ with $G/\gamma_{\mathrm{m}}=16$ are plotted in (a) and (b), respectively. (c) Phase $\theta$ and (d) group delay $\tau$ versus detuning $\Delta^{^{\prime}}$ for $\mathcal{PT}$-symmetric optomechanical systems with three different coupling strengths $G/\gamma_{\mathrm{m}}=(15.1,15.4,15.8)$. Here, $\gamma/\gamma_{\mathrm{m}}=1$ and $\eta=0.3$ are assumed. (e) Group delay $\tau$ (at $\Delta^{^{\prime}}=0$) versus $G$ for different mechanical gains $\gamma/\gamma_{\mathrm{m}}=(1,4,9)$. (f) Group delay $\tau$ versus detuning $\Delta^{^{\prime}}$ for different coupling parameters $\eta=(0.1,0.3,0.6$ and $0.9)$ with $G$/$\gamma_{\mathrm{m}}=15.1$ and $\gamma/\gamma_{\mathrm{m}}=1$. The other system parameters are $\gamma_{\mathrm{m}}=1$ MHz, and $\kappa/\gamma_{\mathrm{m}}$=900. We use word ``Passive-passive" (``Passive-active") to represent conventional optomechanical systems ($\mathcal{PT}$-symmetric-like optomechanical system), respectively. Parameter $\gamma_{\mathrm{m}}$ is the mechanical dissipation rate for conventional optomechanical systems.}%
\label{fig10}%
\end{figure*}

To clearly see the strong enhancement of $T$ and $\Lambda$ around the critical point, the transmission coefficient $T$ and fast dispersion variation $\Lambda$ versus coupling strength $G$ are shown in Fig.~\ref{fig9}(c) and Fig.~\ref{fig9}(d), respectively. Here, the detuning $\Delta^{^{\prime}}$ is set to zero so that the enhancement is strongest. As shown in Fig. \ref{fig9}(c), with $G$ increasing and exceeding the critical point ($G_{\mathrm{cp}}/\gamma_{\mathrm{m}}=15$), the value of $T$ corresponding to the conventional case (see the black-dashed curve) is gradually approaching one, which means the system always exhibits the OMIT phenomenon. However, for the $\mathcal{PT}$-symmetric-like optomechanical system (see the red-solid curve), with $G$ increasing, the system first behaves like the conventional case showing OMIT in the parameter regime $G/\gamma_{\mathrm{m}}<11$. But, as $G$ approaches the vicinity of the critical point, a significant enhancement is observed in the transmission coefficient, which means that electromagnetically-induced amplification (EIA) occurs. As shown in Fig.~\ref{fig9}(d), a strong enhancement for $\Lambda$ is also observed at the vicinity of the critical point [see the red-solid curve]. Such enhancement for $T$ and $\Lambda$ can be easily understood by setting $\gamma\rightarrow\gamma_{\mathrm{m}}$ in Eq.~(\ref{AmplificationAA}). The critical point appears when the denominator of Eq.~(\ref{AmplificationAA}) becomes zero. Physically, such strong enhancement of $T$ and $\Lambda$ originates from the small mechanical-gain-induced strong photoelastic scattering and anti-Stokes field. In other words, the optomechanical interaction and the anti-Stokes field are greatly enhanced around the critical point in our $\mathcal{PT}$-symmetric-like optomechanical system. It is also noted that in our system the control field is always red-detuned.

\subsubsection{Ultra-long group delay}

We now show how the enhanced optical transmission coefficients $T$ and the dispersion variation $\Lambda$ can be used to control the optical
transmission. It is well known that the optical transmission in an EIT window experiences a dramatic reduction in its group velocity, which is also true for the light transmitted in the OMIT window for conventional optomechanical systems~\cite{peit1,peit3,peit4,peit5,peit6}. We now investigate the group delay of the optical signal in our $\mathcal{PT}$-symmetric-like optomechanical system. The optical group
delay of the transmitted light is defined as%
\begin{equation}
\tau=\frac{d\theta}{d\omega_{\mathrm{p}}}.
\label{Timedelay}%
\end{equation}
where $\theta=\arg[t_{\mathrm{p}}(\omega_{\mathrm{p}})]$ is the phase of the output field at the frequency $\omega_{\mathrm{p}}$. The phase $\theta$ and group delay $\tau$ versus detuning $\Delta^{^{\prime}}$ for conventional optomechanical system and the $\mathcal{PT}$-symmetric-like optomechanical system are plotted in Fig.~\ref{fig10}(a) and (b), where an optomechanical coupling strength of
$G/\gamma_{\mathrm{m}}=16$ is assumed. It is clearly shown that the tiny mechanical gain leads to faster variation of the phase [see the red-solid curve in Fig.~\ref{fig10}(a)]. For conventional optomechanical systems [see the black-dashed curve in Fig.~\ref{fig10}(b)],
the maximum group delay is $0.4$ $\mu\mathrm{s}$, which can be prolonged to $14$ $\mu\mathrm{s}$ with mechanical gain [see the red-solid curve]. Furthermore, as shown in Fig.~\ref{fig10}(c), such mechanical can lead to a much faster variation of the phase by decreasing the coupling strength $G$. Also, as shown in Fig.~\ref{fig10}(d), the group delay can be further extended under a weaker coupling strength $G$.

To obtain the optimum group delay (maximum $\tau$), the changes of the group delay $\tau$ versus coupling strength $G$ with fixed detuning $\Delta^{^{\prime}}=0$, are plotted and shown in Figure~\ref{fig10}(e). Three cases with different mechanical gain, [i.e., $\gamma/\gamma_{\mathrm{m}}=(1,4,9)$], are considered. The optimum group delay is always located at the critical point $G_{\mathrm{cp}}=\sqrt{\kappa\gamma}/2$. Compared to the group delay of conventional optomechanical system [see the black-dashed curve in Fig.\ref{fig10}~(b))], a huge enhancement by four orders of magnitude of $\tau$ is obtained around such critical point [see the black-circle curve in Fig.\ref{fig10}~(e))]. Figure~\ref{fig10}(e) also shows that the position of the critical point can be controlled by adjusting the mechanical gain.

In conventional optomechanical systems, the OMIT and the group delay are sensitive to the coupling parameter $\eta$ between the incoming optical field and the cavity field mode. When $\eta$ is set to weak coupling (e.g., $\eta=0.1,0.3$) and overcoupling (e.g. $\eta=0.6,0.9$), the corresponding time delays are $\tau=0.11,0.43,1.48,$ and $8.47$ ($\mu\mathrm{s}$), respectively. Due to environmental disturbances, the coupling may vary (i.e., $\eta$ may change between zero and one). Thus, conventional optomechanical systems are not robust against the change of coupling parameter $\eta$. However, as shown in Fig.~\ref{fig10}(f), for our $\mathcal{PT}$-symmetric-like optomechanical system, the group delay $\tau$ is not very sensitive to the variations in the coupling parameter. These findings open a new direction for controlling the transmission of light beyond what is possible in conventional optomechanical systems~\cite{peit1,peit3,peit4,peit5,peit6}.

\section{Conclusions}

In summary, we have theoretically studied a $\mathcal{PT}$-symmetric like optomechanical system consisting of a mechanical mode with gain and a lossy optical cavity mode. By varying the controllable optomechanical coupling strength $G$, a phase transition is observed for both balanced and unbalanced gain-to-loss cases. We theoretically show that:

(i) for the balanced gain-to-loss case, an amplification window between two Autler-Townes absorption dips appears. The largest amplification is located at the $\mathcal{PT}$ phase transition point. Additionally, by changing the coupling strength $G$, a perfect optical absorbtion is also obtained. The transition from optical absorption (amplification) to amplification (absorption) is controlled by increasing (decreasing) the optomechanical coupling strength $G$;

(ii) for the unbalanced gain-to-loss case and even with a tiny mechanical gain, a huge enhancement of the height of the OMIT-like window and the dispersion variation $\Lambda$ are obtained near the critical point in the broken $\mathcal{PT}$-symmetry regime.

(iii) the group delay $\tau$ of the signal in $\mathcal{PT}$-symmetric-like optomechanical system can be optimized by tuning the optomechanical coupling strength $G$ and improved up to several orders of magnitude ($\tau\sim2$ $\mathrm{ms}$) compared to that of conventional optomechanical system ($\tau\sim1$ $\mu\mathrm{s}$). The maximum group delay is located at the critical point, which can be controlled by the value of mechanical gain.

(iv) the group delay becomes much more robust to the change of the coupling strength $\eta$ between the incoming fields and the optical cavity mode. This can support a stable optical delay.

Note that coherent conversions between microwave and optical fields were demonstrated in a recent experiment~\cite{Andrews}. The mechanical gain can be introduced by a blue-detuned optical pump field. For optomechanical microwave cavities, a controllable coupling between the passive cavity mode and the active mechanical resonator can be achieved with the help of a strong microwave control field (i.e., the control field in our system). So, the $\mathcal{PT}$-symmetric-like optomechanical system is also realizable in the microwave regime for electromechanics. Our theoretical model discussed here can also be generalized to microwave $\mathcal{PT}$-symmetric-like optomechanical system.

These features enable new applications which are hard to realize in conventional optomechanical systems. Our study shows that this $\mathcal{PT}$-symmetric-like optomechanical system could be a powerful platform to control optical signal transmission for $\mathcal{PT}$-symmetric structures.

\section{Acknowledgements}

Y.X.L. is supported by the National Basic Research Program of China 973 Program under Grant No.~2014CB921401, the National Natural Science Foundation of China under Grant Nos.~61328502, 61025022, the Tsinghua University Initiative Scientific Research Program, and the Tsinghua National Laboratory for Information Science and Technology (TNList) Cross-discipline Foundation. R.B.W. acknowledges supports from National nature science Foundation of China under Grant Nos. 61374091,61134008.~\c{S}.~K.~\"{O}. and L.Y. are supported by the Army Research Office under grant No. W911NF-16-1-0339. F.N. was partially supported by the RIKEN iTHES Project, the MURI Center for Dynamic Magneto-Optics via the AFOSR award number FA9550-14-1-0040, the IMPACT program of JST, a Grant-in-Aid for Scientific Research (A), and a grant from the John Templeton Foundation.

\appendix

\section{methods to obtain the mechanical gain}

In this appendix, we will show three different ways to produce mechanical gain [e.g., used in Eq.~(\ref{Lb})], and estimate an upper bound of
the mechanical gain with experimentally accessible parameter values.

\subsection{Mechanical gain by directly driving the mechanical modes}

Experimentally, the mechanical resonator can be coherently driven using either a piezoelectric pump~\cite{phonon,phononpump1}, Josephson phase qubits~\cite{phononpump2}, or a microwave electrical driving~\cite{phononpump3}. A cascaded optical transparency was recently observed by applying a mechanical driving to an optomechanical system with both the optomechanical coupling and parametric phonon-phonon coupling~\cite{phonon1}. Note that the phonon laser is also achieved in an electromechanical resonator by coherently driving the separated mechanical modes~\cite{phonon2,phonon3}. Such mechanical resonator now works as a phonon cavity~\cite{phonon2} which supports high-quality factor mechanical modes with different frequencies. We assume that the pumped mechanical mode $b_{\mathrm{l}}$ is a low-$Q$ one which acts as a phonon cavity with frequency $\omega_{\mathrm{m}^{^{\prime}}}$ and decay rate $\gamma_{\mathrm{m}^{^{\prime}}}.$ Another mechanical mode $b$ is a high-$Q$ one with frequency $\omega_{\mathrm{m}}$ and decay rate $\gamma_{\mathrm{m}}$. The motion of the first mode induces tension that can modify the frequency of the phonon cavity, which introduces a parametric intermodal coupling with strength $G_{\mathrm{ph}}$. The phonon cavity is coherently driven by a blue-detuned mechanical pump $\varepsilon_{\mathrm{m}}\exp^{-i\omega_{\mathrm{m}}t}$ with amplitude $\varepsilon_{\mathrm{m}}$ and frequency $\omega_{\mathrm{m}}$. With the help of the phonon cavity and intermodal coupling, a mechanical gain $\gamma$ is achieved for the high-$Q$ mechanical mode which is given by%
\begin{equation}
\gamma=4\left\vert G_{\mathrm{ph}}\right\vert ^{2}/\gamma_{\mathrm{m}%
^{^{\prime}}}.
\end{equation}

\subsection{Mechanical gain by the phonon lasing method}

In a single cavity, we can find two cavity modes with different frequencies $\omega_{1}$ and $\omega_{2}$, respectively. Here, the frequency-difference of these modes is equal to the frequency of the mechanical resonator (i.e., $\omega_{1}-\omega_{2}=\omega_{\mathrm{m}}$). When pump the high-frequency optical mode is pumped, such two optical modes exchange energy with the mechanical mode through photoelastic scattering~\cite{bahl,chunhua}. The interaction Hamiltonian is given by%
\begin{equation}
H_{\mathrm{m}}=\omega_{1}a_{1}^{\dag}a_{1}+\omega_{2}a_{2}^{\dag}a_{2}%
+\omega_{\mathrm{m}}b^{\dag}b+\beta(a_{1}^{\dag}a_{2}b+b^{\dag}a_{2}^{\dag
}a_{1}),
\end{equation}
where $\omega_{1,2}$ and $\omega_{\mathrm{m}}$ are the frequency of the optical modes and the mechanical mode, respectively. Also, $a_{1,2}$
($a_{1,2}^{\dag}$) are the annihilation (creation) operators of the optical modes; $b (b^{\dag})$ is the annihilation (creation) operator the mechanical mode; and $\beta$ is the coupling coefficient accounting for the modal overlap and scatter gain in the cavity material. Here, the two optical modes can be seen as a two-level atom which exchanges energy with the mechanical phonons. To clearly see this, we define the ladder and population inversion operators as%
\begin{align}
J_{+}  &  =a_{1}^{\dag}a_{2},\\
J_{-}  &  =a_{2}^{\dag}a_{1},\\
\mathcal{N}  &  \mathcal{=}a_{1}^{\dag}a_{1}-a_{2}^{\dag}a_{2}.
\end{align}
The Heisenberg equations of motion for the mechanical mode $b$ and the operator $J_{-}$ (with damping added) are%
\begin{align}
\dot{b}  &  =-\left(  \gamma_{\mathrm{m}}+i\omega_{\mathrm{m}}\right)
b-i\beta J_{-},\label{phonong1}\\
\dot{J}_{-}  &  =-\left(  \gamma_{\mathrm{a}}+i\Delta_{\omega}\right)
J_{-}+i\beta\mathcal{N}b, \label{phonong2}%
\end{align}
where $\Delta_{\omega}=\omega_{1}-\omega_{2}.$ $\gamma_{\mathrm{m}}$ is the decay rate of the mechanical mode; Here, $\gamma_{\mathrm{a}}=\left(  \gamma_{\mathrm{1}}+\gamma_{\mathrm{2}}\right)  /2,$ where $\gamma_{\mathrm{1}}$ and $\gamma_{\mathrm{2}}$ are the decay rate of the optical modes $a_{1}$ and $a_{2}$, respectively. Next, we introduce the rotating frame by setting
$B=\exp(i\omega_{\mathrm{m}}t)b$ and $J=\exp(i\omega_{\mathrm{m}}t)J_{-}$, in which Eqs.~(\ref{phonong1}) and ~(\ref{phonong2}) can be rewritten as%
\begin{align}
\dot{B}  &  =-\gamma_{\mathrm{m}}B-i\beta J,\label{shibianb}\\
\dot{J}  &  =-\left[  \gamma_{\mathrm{a}}+i\left(  \Delta_{\omega}%
-\omega_{\mathrm{m}}\right)  \right]  J+i\beta\mathcal{N}B.
\end{align}
Because the coupling between the optical mode and the mechanical mode is weak, and the cavity decay rate is much larger than the mechanical dissipation rate [i.e., $\gamma_{\mathrm{a}}\gg(\beta,\gamma_{\mathrm{m}})$ ], we can adiabatically eliminate the optical mode $J$ by setting $\dot{J}=0$, which leads to%
\begin{equation}
J=\frac{i\beta\mathcal{N}}{\gamma_{\mathrm{a}}+i\left(  \Delta_{\omega}%
-\omega_{\mathrm{m}}\right)  }B, \label{wentaij}%
\end{equation}
Substituting Eq.~(\ref{wentaij}) into Eq.~(\ref{shibianb}), we find%
\begin{equation}
\dot{B}=\left[  \frac{\beta^{2}\mathcal{N}}{\gamma_{\mathrm{a}}+i\left(
\Delta_{\omega}-\omega_{\mathrm{m}}\right)  }-\gamma_{\mathrm{m}}\right]  B,
\end{equation}
which indicates an effective mechanical gain%
\begin{equation}
\gamma=\frac{\beta^{2}\mathcal{N}\gamma_{\mathrm{a}}}{\gamma_{\mathrm{a}}%
^{2}+\left(  \Delta_{\omega}-\omega_{\mathrm{m}}\right)  ^{2}}. \label{Gain}%
\end{equation}
Select the two optical modes so that $\Delta_{\omega}=\omega_{1}-\omega
_{2}=\omega_{\mathrm{m}}$, and then Eq.~(\ref{Gain}) can be simplified to%
\begin{equation}
\gamma=\frac{G^{2}}{\gamma_{\mathrm{a}}}. \label{Glast}%
\end{equation}
where $G=\sqrt{\mathcal{N}}\beta$ is the control-field enhanced optomechanical coupling rate. Under the strong-pump condition, a remarkable mechanical gain can be achieved. For example choosing $G/\gamma_{\mathrm{a}}=5$ leads to a mechanical gain of $\gamma/\gamma_{\mathrm{a}}=25$.

\subsection{Mechanical gain by blue-detuning optical pump}

For an optomechanical system with a strong blue-detuned control field e.g., $\Delta=\omega_{\mathrm{m}}$, the anti-Stokes parameter process is greatly enhanced in contrast to the greatly-suppressed Stokes parameter process. An effective gain coefficient for the mechanical mode is given by $C=4\left\vert G\right\vert ^{2}/\kappa\gamma_{\mathrm{m}}$. Indeed phonon amplification has been experimentally observed in various optomechanical systems~\cite{pam1,pam3,pam4,pam5,pam6,pam7}.

\end{document}